\documentclass[final,a4paper]{siamart190516}

\usepackage{graphicx}
\usepackage{caption}
\usepackage{subcaption}
\usepackage[utf8]{inputenc}

\usepackage{amsmath,amsfonts,amssymb,bm,relsize,tikz}

\usepackage{bef_alex_nothm}

\newsiamremark{remark}{Remark}

\usetikzlibrary{arrows.meta}

\usepackage[notref,notcite,color]{showkeys}

\newcommand{\DLL}{\Lambda^*}   
\renewcommand{\ti}{{\times}}
\renewcommand{\dot}[1]{\overset{\text{\LARGE.}}{#1}}
\renewcommand{\ddot}[1]{\overset{\text{\LARGE.\!\:\!.}}{#1}}
\newcommand{\far}{\mafo{far}}
\newcommand{\DHW}{{\upshape DHW}}  
\newcommand{\Ew}{\mafo{Ew}}  
\newcommand{\whg}{{\wh{g}}}  
\newcommand{\mfH}{\mathfrak H}
\newcommand{\mfS}{\mathfrak S}
\newcommand{\STEP}[1]{\noindent{\emph{#1}}} 
\newcommand{\mathOP}[1]{\mathop{\mathrm{#1}}} 
\newcommand{\NNeins}[1]{{\wh c_\bbU^{#1}}} 

\usepackage{graphicx}
\graphicspath{{figs/}}

\headers{On the Darwin--Howie--Whelan equations}{Th. Koprucki, A. Maltsi, and A. Mielke}

\title{On the
  Darwin--Howie--Whelan equations for \\ 
  the scattering of fast electrons described by \\ the Schr\"odinger equation%
  \thanks{Submitted to SIAP January 13, 2021. Accepted April 22,
    2020.%
    \funding{Research partially funded by Deutsche Forschungsgemeinschaft 
    via Berlin Mathematics Research Center MATH+ (EXC-2046/1, project
    ID: 390685689), subprojects EF3-1 and AA2-5.}}}
 
\author{Thomas Koprucki\thanks{Weierstraß-Institut f\"ur Angewandte
   Analysis und Stochastik, 10117 Berlin, Germany.}
  \and
        Anieza Maltsi\footnotemark[2]
\and
        Alexander Mielke\footnotemark[2]
  \thanks{Humboldt
 Universit\"at zu Berlin, Institut f\"ur Mathematik, 12489 Berlin,
 Germany} 
}

\date{Submitted January 13, 2021. Accepted  April 22, 2021} 

\begin{document}
 
\maketitle

\begin{abstract}
The Darwin--Howie--Whelan equations are commonly used to describe and simulate
the scattering of fast electrons in transmission electron microscopy. They are
a system of infinitely many envelope functions, derived from the Schr\"odinger
equation. However, for the simulation of images only a finite set of envelope
functions is used, leading to a system of ordinary differential equations 
in thickness direction of the specimen. We study the mathematical
structure of this system and provide error estimates to evaluate the accuracy
of special approximations, like the two-beam and the systematic-row
approximation.
\end{abstract}





\section{Introduction}
\label{se:intro}

The Darwin--Howie--Whelan (DHW) equations, which are often simply called
Howie--Whelan equations (cf.\ \cite[Sec.\,2.3.2]{Jame90APTH} or
\cite[Sec.\,6.3]{Kirk20ACEM}), are widely used for the numerical simulation of
transmission-electron microscopy (TEM) images, e.g. see \cite{Nier19pyTEM} for
the software package \texttt{pyTEM} or \cite{SchSta93,WuSch19TEMD} for the
software \texttt{CUFOUR}. They describe the propagation of electron beams
through crystals and can be applied to semiconductor nanostructures, see
\cite{Degr03ICTE, Pascal2018, MNBL19DDEG, MNSTK20NSTI}.  They provide a theoretical basis
that allows one to construct suitable experimental set ups for obtaining
microscopy data on the one hand, and can be used to analyze measured data in
more details on the other hand.  The origins of this model go back to Darwin in
\cite{Darw14TXRR12} with major generalizations by Howie and Whelan in
\cite{HowWhe61DCEM}. Moreover, the DHW equations are closely related to the
approach based on the Bethe potentials used in \cite{WanDeg15MDES}.  

Currently, many quantitative methods emerge for applications in TEM
\cite{Nier19pyTEM,WuSch19TEMD}, holography \cite{LJCRGH14DSTD,JLCRCHGH14DESM},
scanning electron microscopy \cite{PSCHTD19EWDS,Pasc19DMND}, electron
backscatter diffraction \cite{WTSDP07MBSE,ZhuDeg20EBSD}, and electron
channelling contrast imaging \cite{Pascal2018}, where quantitative
evaluations of micrographs are compared to simulation results to replace former
qualitative observations by rigorous measurements of embedded structures in
crystals.  For that reason it is essential to evaluate the accuracy and the
validity regime of the chosen modeling schemes and simulation tools. 
In electron microscopy this includes the heuristic approaches to select the relevant
beams in multi-beam approaches \cite{WTSDP07MBSE,Nier19pyTEM,WuSch19TEMD}.  
The present work is devoted to the theory behind the DHW equations and thus
provides mathematical arguments and refinements for the beam-selection problem. 
 
The DHW equations can be derived from the time-dependent Schr\"odinger equation
for the wave function $\psi(t,x)$ of the electrons:  
\begin{equation}
  \label{eq:I.TDSE}
  \ii \hbar \,\fr{\partial\psi(t,x)}{\partial t} = - \fr{\hbar^2}{2m} \,
  \Delta \psi(t,x) - q V_\rmC(x) \psi(t,x),
\end{equation}
such that $|\psi(t,x)|^2$ denotes the probability density of the
electrons.  Here $x$ denotes the position in the specimen, $V_\rmC$ is a
periodic potential describing the electronic properties of the crystal,
$m=m_0\gamma$, with $\gamma$ the relativistic mass ratio and $m_0$ the electron
rest mass, $\hbar$ Planck's constant and $q$ the elementary charge.  Using
$  
  \psi(t,x) = \ee^{-\ii 4\pi^2\fr{\hbar}{2m}|k_0|^2t} \Psi(x)
$  
we obtain the static Schr\"odinger equation
\begin{equation}
  \label{eq:SSE}
  \Delta \Psi(x) + (2\pi |k_0|)^2 \Psi(x)  = -4\pi^2\calU(x) \Psi(x), 
\end{equation}
where $k_0$ is the wave vector of the incoming beam and $\calU$ is the reduced
electrostatic potential defined as $\calU(x) = \fr{2 m q}{\hbar^2} V_\rmC(x)
$. The periodicity lattice of the crystal and its potential $\calU$ is
denoted by $\Lambda$ and its 
dual lattice by $\Lambda^*$.  

\begin{figure}
\centerline{
\begin{tikzpicture}[scale=0.8]
\draw[ thick, color=gray,  ->] (-1.8,2)--(-1.8,-1.5) node[left]{$z$};
\draw[ thick, color=gray, ->] (-2.4,1)--(3.5,1) node[right]{$y$};
\node[left] at (-2.45,1){$z=0$};
\node[left] at (-2.3,-0.4){$z=z_*$};
\draw[  color=blue, very thick, fill = blue!30] (-2,-0.4)--(3,-.4)
 node[right]{specimen}--(3,1)--(-2,1)--cycle;
\draw[color=red, very thick] 
\foreach \x in {0,...,6}
{(1.4+0.05*\x,1.4+0.2*\x)--(2.2+0.05*\x,1.2+0.2*\x)};
\draw[color=red!70!black, ultra thick, ->] 
      ( 2.1,2.7 ) node[right]{\!$k_0$}--(1.7,1.1);
\draw[color=red, thick] 
\foreach \x in {-2,...,4}
{(0.6+0.05*\x,-1.6+0.2*\x)--(1.4+0.05*\x,-1.8+0.2*\x)};
\draw[color=red!70!black, very thick, ->] 
      ( 1.25,-0.7 )--(0.85,-2.3) node[left]{\!$k_0$};
\draw[color=red!50!blue, thick] 
\foreach \x in {-2,...,4}
{(1.4+0.01*\x,-1.63+0.2*\x)--(2.2+0.01*\x,-1.7+0.2*\x)};
\draw[color=red!50!blue, very thick, ->] 
      ( 1.85,-.5 )--(1.7,-2.5) node[right]{\!$k_0{+}g$};
\begin{scope}
\clip  (-1.95,-0.35) rectangle (2.95,0.95) ;
 \foreach \i in {-10,...,30}
         \foreach \j in {-2,...,20}
          \fill (-2.1 +0.2*\i+0.1*\j , 0.2*\j-0.05*\i) circle(0.03);
\end{scope}
\draw[very thick, ->] (-0.8,-0.5) -- (-0.8,-1.5) node[below]{$\bfnu$};
\end{tikzpicture}
\quad
\begin{minipage}[b]{0.32\textwidth}
\caption{The incoming wave with wave vector $k_0$ enters the specimen,
  is partially transmitted, and generates waves with nearby wave
  vectors $k_0{+}g$.}
\label{fig:ExperSetup}
\end{minipage}
}
\end{figure}
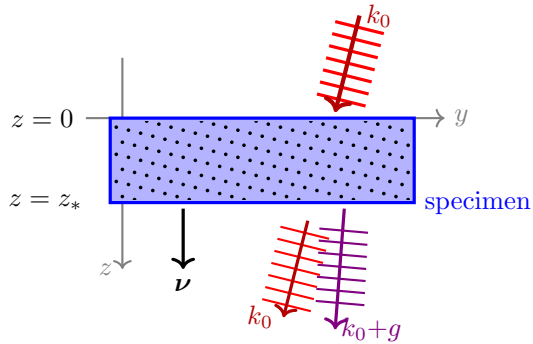

We decompose the spatial variable $x$ into the transversal part $y$ orthogonal
to the thickness variable $z\in [0,z_*]$, where $z=0$ is the side where the
monochromatic electron beam
$\psi(t,x)= \ee^{\ii (-4\pi^2\fr{\hbar}{2m}|k_0|^2t +2\pi k_0 \cdot x)} $ with
wave vector $k_0$ enters, and $z=z_*$ is the side where the scattered beam
exits the specimen, see Figure \ref{fig:ExperSetup}.  
The so-called ``column approximation'' restricts the focus to solutions of
\eqref{eq:SSE} that are exactly periodic in $y$ and are slow modulations in $z$
of a periodic profile in $z$. Hence, we seek solutions in the
form
\begin{equation}
\label{eq:I.ColumnAppr}
  \Psi(x) = \sum_{g\in \DLL} \psi_g(z) \,\ee^{\ii\,2\pi k_0\cdot
    x}\,  \ee^{\ii\,2\pi  g\cdot x},
\end{equation}
where $\DLL\subset \R^d $ denotes the dual lattice and $\psi_g$ is the
slowly varying envelope function of the beams in the directions of the
wave vector $g \in \DLL$. Note that now $g=0$ corresponds to the main
incoming beam. 
Inserting \eqref{eq:I.ColumnAppr} into \eqref{eq:SSE} and dropping the
term $\frac{\rmd^2}{\rmd z^2}\psi_g$ one obtains the DHW
equations for infinitely many beams  (see Section \ref{se:Model} for more
details on the modeling): 
\begin{align}
  \label{eq:I.DHW}
&  \frac{\rho_g}\pi \, \frac{\rmd}{\rmd z} \psi_g(z) = \ii \Big(\sigma_g
  \psi_g(z) + \sum_{h\in \DLL} U_{g-h}\psi_h(z) \Big), \quad
  \psi_g(0)=\delta_{0,g}, \quad \text{for } g\in \DLL \\
& \text{where } \ \rho_g = (k_0{+}g)\cdot \nu \ \text{ and }
   \ \sigma_g= |k_0|^2- |k_0{+}g|^2, \nonumber
\end{align}
where $\nu=(0,...,0,1)^\top$ is the normal to the crystal surface and where
$U_g$ are the Fourier coefficients of the periodic potential $\calU$, i.e.\
$\calU(x) = \sum_{g\in \DLL} U_g \ee^{\ii 2\pi g\cdot x}$.

In fact, to make \eqref{eq:I.DHW} equivalent to the full static
Schr\"odinger equation \eqref{eq:SSE} one has to add the second
derivatives with respect to $z$, namely
$\frac1{4\pi^2} \frac{\rmd^2}{\rmd z^2} \psi_g(z)$. Dropping these terms 
constitutes the DHW equations, which are solved as an initial-value problem
with the simple initial condition $\psi_g(0)=\delta_{0,g}$ (Kronecker
symbol) for the incoming beam, and $\big(\psi_g(z_*)\big)_{g\in \Lambda^*}$
describes all exiting beams.  In contrast, the full second-order
equations would need a careful setup of transmission and reflection conditions
at $z=0$ and $z=z_*$, and then are able to account for the backscattering of
electrons.  We refer to \cite{vDyc76IBHE} and our Remark \ref{re:JustifyDrop}
for the justification of this approximation for electrons with high energy that
are typical for TEM.

Most often, the DHW equations are stated in the form that the equation for
$\psi_g$ is divided by $\rho_g/\pi$, and then it features the important
\emph{excitation error} $s_g:= \sigma_g/(2\rho_g)$.  However, the mathematical
structure can be seen better in the form \eqref{eq:I.DHW}: the right-hand side
is given by the imaginary unit $\ii$ multiplied by a Hermitian operator 
under our standard assumption that $\calU$ is a real potential, i.e.\
$U_{-g}=\ol U_g$. This will be crucial for the subsequent analysis 
based on the associated Hamiltonian structure. 

In the physical literature, the DHW equations are formally stated as a system
for infinitely many beam amplitudes $\psi_g$ with $g$ running through the whole
dual lattice $\DLL$.  For the numerical solution one has to select a
finite set $\bfG\subset \DLL$ of relevant beams, e.g. the classical
two-beam approximation, see Section \ref{su:TwoBeam}. To our knowledge
there is no systematic discussion about the accuracy of approximations
depending on the choice of $\bfG$.  The main goal of
this paper is to provide mathematical guidelines for optimal choices that are
justified by exact error estimates.

First, we observe that \eqref{eq:I.DHW} for \emph{all} $g\in \DLL$ is probably
ill-posed, in particular, because of $\rho_g$ changing sign and, even worse,
becoming $0$ or arbitrarily close to $0$.  It is clear that neglecting the term
$\frac1{4\pi^2} \frac{\rmd^2}{\rmd z^2} \psi_g(z)$ cannot be justified for such
$g$'s.  Hence, one should realize that the DHW equations \eqref{eq:I.DHW} is
only useful for $g$ where $\rho_g$ is close to $\rho_0=k_0\cdot \nu>0$, 
see Section \ref{su:RelevaWaveVect}. 
But the main questions of beam selection remain:
\\
\textbullet\  What does ``close'' mean? \\
\textbullet\   How many and which beams are needed to
obtain a reliable approximation for the\\ 
\phantom{\textbullet\ }solution of the Schr\"odinger equation,
in particular for high-energy electron beams?\smallskip 

We approach these questions by systematically investigating the dependence of
the solutions $\bfpsi^\bfG=(\psi_g)_{g\in \bfG}$ on the chosen subset $\bfG$ of
the dual lattice $\DLL$ for which we solve \eqref{eq:I.DHW}. More precisely,
for $\bfG\subset \DLL$ we define \DHW$_\bfG$ to be the set of equations
\begin{align}
  \label{eq:I.DHW.bfG}
  &  \frac{\rho_g}\pi \, \frac{\rmd}{\rmd z} \psi_g(z) = \ii \Big(\sigma_g
  \psi_g(z) + \sum_{h\in \bfG} U_{g-h}\psi_h(z) \Big), \quad
  \psi_g(0)=\delta_{0,g}, \quad g\in \bfG.
\end{align}
We will shortly write this in vector-matrix form 
\begin{align}
  \label{eq:I.DHW.bfG2}
&  R \dot \bfpsi(z) = \ii \big(\Sigma
   {+} \bbU\big)\, \bfpsi(z) , \quad
  \bfpsi(0)= (\delta_{0,g})_{g\in \DLL}, \quad \bfpsi=(\psi_g)_{g\in \bfG} .
\end{align}
For $\gamma \in {]0,1[}$ and $M>0$, we define two important classes of
\emph{admissible beam sets} by
\[
  \bfG_\gamma :=\bigset{g\in \DLL}{ \rho_g \geq \gamma \rho_0} \quad \text{and}
  \quad \bfG^M := \bigset{g \in \DLL}{ |g|\leq M},
\]
 such that always $0 \in \bfG^M \cap \bfG_\gamma$ and $\rho_g >0$
for $g \in \bfG_\gamma$.  Throughout we will only consider the case of such $M>0$
that $\bfG^M \subset \bfG_\gamma$ for some $\gamma>0$.

We first show in Proposition \ref{pr:ExUniFlux} that for each
$\bfG\subset \bfG_\gamma$ the system \DHW$_\bfG$ has a unique solution
$\bfpsi^\bfG:\R\to \mfH(\bfG)$, where $\mfH(\bfG)$ is the Hilbert space
generated by the scalar product
\[
  \big\langle \bfpsi, \bfvarphi\big\rangle_\bfG := \sum_{g\in \bfG} \rho_g
  \psi_g \ol \bfvarphi_ g \quad \text{and the norm } \|\bfpsi\|_\bfG:=
  \big\langle \bfpsi, \bfpsi\big\rangle_\bfG^{1/2}.
\] 
In Section \ref{su:CutOffError} we will show that the influence of the exact
choice of the set $\bfG$ is not important if we stay inside $\bfG_\gamma$ and
if we have enough modes around $g=0$.  More precisely, for two sets
$\bfG^{(1)}$ and $\bfG^{(2)}$ satisfying
$\bfG^M \subset \bfG^{(j)} \subset \bfG_\gamma$ the unique solutions
$\bfpsi^{(j)}$ can be compared on $\bfG^M$ as follows: 
\begin{equation}
  \label{eq:I.est}
 \big\| \bfpsi^{(1)}(z)|_{\bfG^M} -  \bfpsi^{(2)}(z)|_{\bfG^M} \big\|_{\bfG^M}
\leq \wh C_\calU\:\ee^{\wh\kappa|z| - \wh\alpha M}\,\|\bfpsi(0)\|_\bfG  \quad
    \text{for all } z\in \R\,. 
\end{equation}
The coefficients $\wh\kappa$ and $\wh\alpha$ can also be given explicitly,
see Corollary \ref{co:ArbiSets}.  For this estimate, we use the
fundamental assumption that the Fourier coefficients $U_g$ of the scattering
potential $\calU$ decay exponentially:
\begin{equation}
  \label{eq:I.Ug.expon}
  |U_g| \leq C_\bbU \,\ee^{-\alpha_\bbU |g|} \quad \text{for all } g\in \DLL. 
\end{equation}
Hence, we see that scattering allows the energy to travel from the transmitted
beam $g=0$ to the diffracted beams linearly with respect to the distance
$|z|$. Note that we interpret equation \eqref{eq:I.DHW.bfG2} as an autonomous
Hamiltonian system, such that estimates like \eqref{eq:I.est} hold for all
$z\in \R$. However, to evaluate realistic errors for a specimen we restrict to
$z\in [0,z_*]$, see e.g.\ \eqref{eq:I.estLOLZ}. In particular, \eqref{eq:I.est}
provides a good bound on $[0,z_*]$ as long as $z_*$ is much smaller than
$\wh\alpha M/\wh \kappa$. 

In a second step we are able to reduce the set $\bfG$ even further by
restricting $g$ into a neighborhood of the Ewald sphere
\[
\bbS_\Ew := \bigset{ g \in \R^d \:}{ \:|k_0|^2 -
|k_0{+}g|^2 =0}.
\] 
Indeed, in TEM the incoming beam with wave vector $k_0$ is chosen exactly in
such a way that the intersection of the Ewald sphere $\bbS_\Ew$ with the dual
lattice $\DLL$ contains, in addition to the transmitted beam $g=0$, a special
number of other points. 

From the energetic point of view it is important to observe that the modulus
$|s_g|=|\sigma_g|/(2\rho_g)$ of the excitation errors for wave vectors $g$ not
close to the Ewald sphere are much bigger than $|U_g|/\rho_g$.  To exploit
this, we use the classical norm and energy conservation for the
 linear Hamiltonian system 
\eqref{eq:I.DHW.bfG2}, namely  $\| \bfpsi(z)\|_\bfG = \|\bfpsi(0)\|_\bfG$
and $\| R^{-1} (\Sigma{+}\bbU) \bfpsi(z)\|_{\bfG} = \| R^{-1} (\Sigma{+}\bbU)
\bfpsi(0)\|_{\bfG}$ together with the estimate
\[
\| R^{-1} \Sigma \bfpsi(z) \|^2_{\bfG} \leq 2 \| R^{-1} (\Sigma{+}\bbU)
\bfpsi(z)\|^2_{\bfG} + 2 \| R^{-1} \bbU \bfpsi(z)\|^2_{\bfG}.
\]
Since $\| R^{-1} (\Sigma{+}\bbU) \bfpsi(0)\|_{\bfG} $ is controlled
by the initial value $\bfpsi(0)=\bfdelta =(\delta_{0,g})$ and
$ \| R^{-1} \bbU \bfpsi(z)\|^2_{\bfG}  \leq \NNeins{} \| \bfpsi(z)\|^2_{\bfG}$, we
obtain a good bound on $\sum_{g\in \bfG} \rho_g |s_g\psi_g(z)|^2$ in terms
of the initial data. This allows us to quantify the smallness of the
amplitudes $|\psi_g(z)|$ if the excitation error $|s_g|$ lies
above a cut-off value $\wt s_*$, see Section~\ref{su:Averaging}.

With this, we provide an error bound for the so-called Laue-zone approximation
$\bfpsi^\mafo{LOLZ} $ (cf.\ Section \ref{su:LOLZ}), 
where we choose $M \sim |k_0|^{1/2}$ to approximate a spherical cap of the
Ewald sphere, which has the height of one dual lattice spacing. 
For the cut-off $\wt s_*$ one can
choose a constant that is proportional to the spacing of the dual lattice. 
The final error bound compares the solutions $\bfpsi^\gamma$ and
$\bfpsi^\mafo{LOLZ} $ of \DHW$_{\bfG_\gamma}$ and \DHW$_{\bfG_\mafo{LOLZ}}$,
respectively, on the interval $z \in [0,z_*]$: 
\begin{equation}
  \label{eq:I.estLOLZ}
  \| \bfpsi^\mafo{LOLZ}(z)- \bfpsi^\gamma(z)|_{\bfG_\mafo{LOLZ}} 
 \|_{\bfG_\mafo{LOLZ}} \leq N_1 \Big( \frac1{|\alpha_* k_0|^2} +
 \frac{\alpha_* C_\bbU^2}{|k_0|^2}  \,z_*\Big)\|\bfdelta\|_\bfG\,.
\end{equation}
Here $N_1$ is a computable, dimensionless constant, and $\alpha_*$ is the lattice
constant of $\Lambda$. The first error term arises from the restriction 
of $\bfG_\gamma$ to $\bfG^M$, and it is small for high energies, i.e.\
$|\alpha_*k_0|\gg 1$. The second error term arises from the restriction to the
neighborhood of the Ewald sphere and has the form 
\[
   \frac{\alpha_*}{\ell_\mafo{scatt}}\:  \frac{z_*}{\ell_\mafo{scatt}} \qquad
\text{with global scattering length }\ell_\mafo{scatt}= \frac{|k_0|}{C_\bbU}.  
\] 
Since we always have $z_*\approx \ell_\mafo{scatt}$ we see that the error is
small if the scattering length $\ell_\mafo{scatt} $ is much bigger than the
lattice constant $\alpha_*$, which is indeed the case in TEM experiments with
specimens of about 100 atomic layer thickness. 

A similar error analysis is then done for the two-beam approximation and the
systematic-row approximation, see Sections \ref{su:TwoBeam} and
\ref{su:SystematicRow}. Finally, Section \ref{se:Applications} presents numerical
simulations that underpin the quality of the error bounds and thus
provide a justification of the numerics done when solving the DHW equations as
in \cite{Nier19pyTEM}.

The authors are not aware of any mathematical analysis for beam-like,
high-energy solutions of the static Schr\"odinger equation.  There is a large
body of works on the semiclassical limit of the time-dependent Schr\"odinger
equation $\ii \eps \pl_t \psi =-\eps^2 \Delta \psi + V(\frac1\eps x)\psi$ with
a periodic potential (cf.\ \cite{HoSpTe01SLSE, BeEsPu04AAQS, JiMaSp11MCMS})
that are related in spirit. There, $|k|$ is of order $1/\eps \gg1$ such that
$\eps|k|\approx 1$ is a different regime as our case $\alpha_*|k_0|\gg 1$.
Moreover, there initial-value problem for the time-dependent Schr\"odinger
equations are studied assuming smooth envelope functions, while we are
concerned with static scattering-type solutions of a Helmholtz-type
Schr\"odinger equation. 
 
We close this introduction with some remarks concerning the usage of the DHW
equations in TEM imaging of objects embedded in crystals, such as quantum dots.
There, the crystal structure changes slightly because of the different
properties of the embedded materials. In the simplest case one assumes that the
crystal lattice stays intact, but the potential $\calU$ is no longer exactly
periodic but varies on a larger length scale.  Nevertheless one can assume that
electron beams can propagate vertically through the crystal, i.e.\ the
column approximation holds. As the embedded material has different chemical
composition the potential remains locally periodic, but on a mesoscale 
depends on the thickness variable $z\in [0,z_*]$ as well as the horizontal
variables $y_1$ and $y_2$.  Using the column approximation the latter variables
are considered as parameters that are used for scanning the probe pixel by
pixel. The associated DHW equations then take the form 
\[
R\frac{\rmd}{\rmd z} \bfpsi = \ii \big( \Sigma + \bbU (y_1,y_2,z) 
\big) \bfpsi, \quad \bfpsi(0)=\bfdelta. 
\]

Typically, the objective aperture of the microscope selects only a single beam
$\whg\neq 0$ (e.g.\ $g=(0,4,0)$ in \cite{MNSTK20NSTI}), and its exit intensity
\begin{figure}
\centering
\includegraphics[width=\textwidth, trim=0 40 0 50, clip=true]{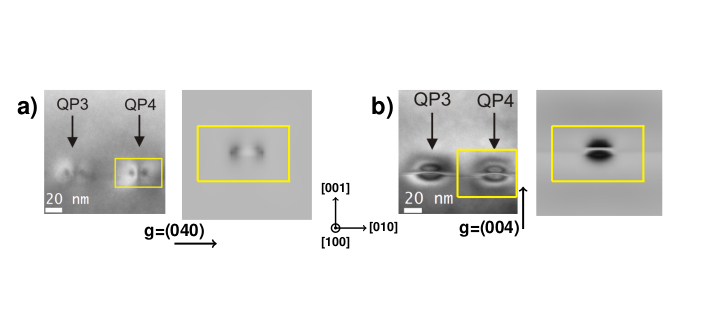}
\caption{TEM images of lens-shaped InAs quantum dots embedded in a GaAs
  matrix: (a) Experimental dark field of
  (040) beam and the corresponding image from a DHW simulation. 
  (b) Experimental dark field of (004) beam and the corresponding image
  from a DHW simulation.  Yellow areas
  indicate the same areas in experimental and simulated images.  Adapted by
  permission from Springer Nature: Optical and Quantum Electronics,
  \cite[Fig.\,6]{MNSTK20NSTI}, $\copyright$ 2020.\label{fig:tem.image}}
\end{figure}
$I_\whg(y_1,y_2):=|\psi_\whg(z_*;y_1,y_2)|^2$ is recorded in dependence of the
horizontal position $(y_1,y_2)$. The analysis of the $(y_1,y_2)$-dependence
goes beyond of this work and will be addressed in future research. In general,
the term ``\emph{bright-field image}'' is used, when the undiffracted beam
related to $\whg=0$ is included in the aperture. The term ``\emph{dark-field}''
is used for images, where only one diffracted beam $\whg \neq 0$ forms the
image. In Figure \ref{fig:tem.image} we see two dark-field TEM images of InAs
quantum dots, for the $(040)$ beam and the $(004)$ beam, and the corresponding
simulated images, all taken from \cite{MNSTK20NSTI}.

The methods developed here for the DHW equations address more generally
the question of beam selection for electron waves in crystals, which is also
important for calculations based on the Bloch wave expansion
\cite{WanDeg15MDES} or electron backscatter diffraction \cite{WTSDP07MBSE}. In
particular, our methods provide mathematical error estimates that allow us to
understand and refine beam selection scheme in situations where the classical
two-beam approach is not sufficient, see \cite{ViMaLa17CCPF, MNBL19DDEG,
  WuSch19TEMD, MNSTK20NSTI}. The problem of beam selection will be even
more important because of the recent trend to use lower acceleration voltages,
see e.g.\ \cite{Pasc19DMND, Pascal2018}, where $|k_0|$ is smaller and
scattering into more beams occurs naturally. This is also why we try to make
all estimates as explicit as possible in their dependence on the data such as
$|k_0|$, $C_\bbU$, and $\alpha_\bbU$ (cf.\ \eqref{eq:I.Ug.expon}).

\section{The modeling}
\label{se:Model}

\subsection{Derivation of the DHW equations}
\label{su:DerivDHW}

Transmission electron microscopy uses high-energy electron beams, which
can be described by the relativistic wave equation for an electron in
an electrostatic field, see \cite{Degr03ICTE}:
\begin{equation}
  \label{eq:Mod.Schr}
  \Delta \Psi(x) + (2\pi |k_0|)^2 \Psi(x)  = -4\pi^2\calU(x) \Psi(x), 
\end{equation}
where $k_0$ is the wave vector of the incoming beam, and $\calU$ is the reduced
electrostatic potential.  The modulus of the wave vector is related to the
(relativistic) wave length by $|k_0| = 1/\lambda_0$. The
wave length $\lambda_0$ is obtained from the acceleration voltage $E$ via
\[
\lambda_0=\hbar\big/{\sqrt{2 m_0 q E \big( 1 {+} \tfrac{q E}{2m_0 c_0^2} \big) }},
\]
where $\hbar$ is the Planck's constant, $m_0$ is the electron rest mass, $q$ is
the elementary charge, and $c_0$ is the speed of light.  The \emph{reduced
  electrostatic potential} is given by
\begin{equation}
   \calU(x) = \fr{2 m_0 q}{\hbar^2}\, \gamma \, V_\rmC(x) \quad 
   \text{with the relativistic mass ratio } \gamma = 1 + \frac{q E}{m_0c_0^2}\,.
\end{equation}
Here $V_\rmC$ is the (possibly complex) electrostatic potential such that the
reduced potential $\calU$ has the unit $\mafo{m}^{-2}$.  The table in
Figure~\ref{fig:physical_quantities} shows typical values for the wave vector
$k_0$ and the mass ratio $\gamma$ for different values of the acceleration
voltage $E$.
\begin{figure}[H]
\centerline{
\begin{minipage}[c]{0.41\textwidth}
\begin{tabular}{|c||c|c|c|} \hline 
$E$  &  $k_0$ & $\gamma$ & $\beta=v/c_0$\\ \hline\hline
100 & 270.165& 1.196 & 0.548  \\ \hline
200 & 398.734  & 1.391 & 0.695 \\ \hline
300 & 507.937 & 1.587 & 0.777   \\ \hline
400 & 608.293 & 1.783 & 0.828 \\ \hline
\end{tabular}
\end{minipage}\qquad
\begin{minipage}[c]{0.44\textwidth}
\caption{\rule{12.1em}{0pt}}\vspace{-0.6em}
 Acceleration voltage $E$ in $\mafo{kV}$,\\ wave number
    $k_0$ in $\mafo{nm^{-1}}$,\\
   mass ratio $\gamma$, and  \\
  relative velocity $\beta=v/c_0$. \\
\small (Table adapted from \cite[p.\,93 Table\,2.2]{Degr03ICTE})%
\label{fig:physical_quantities}%
\end{minipage}
}
\end{figure} 

The periodicity of the potential $\calU$ is given by the (primal) 
lattice $\Lambda \subset \R^d$ via $\calU(x{+}r) = \calU (x)$
for  all $x\in \R^d$ and all lattice vectors $r \in \Lambda$.  The dual lattice is
\[
  \DLL:= \bigset{g \in \R^d}{ g\cdot r \in \Z \text{ for all
    } r\in \Lambda }. 
\]
With this, we are able to write $\calU$ by its Fourier expansion 
$ 
\calU (x) = \sum_{g\in \DLL} \ee^{\ii\,2\pi  g\cdot x} \: U_g. 
$ 

The solution $\Psi$ of the Schr\"odinger equation is assumed to have an
envelope form given by a plane wave $\ee^{\ii\,2\pi k_0\cdot x}$ times a
slowly varying function $\wt \Psi$, where $k_0$ is the
wave vector for the incoming electron beam.
Throughout the paper, we decompose $x\in \R^d$ into a in-plane component $y \in \R^{d-1}$
and a transversal component $z\in \R$, i.e.\ after rotating the coordinate axis
we have $x=(y,z)$. To comply with physicists convention, the $z$ direction is
orientated roughly parallel to the electron beam, while the outwards normal to the
specimen at the exit plane $z=z_*$ denoted by $\bfnu$, 
is assumed to be
\[
\bfnu := (0,\ldots,0,1)^\top.
\]
We emphasize that the lattices $\Lambda$ and $\DLL$ are not necessarily aligned
with one of the directions $\bfnu$ or $k_0$, but we always assume $k_0\cdot
\bfnu >0$, see Figure \ref{fig:ExperSetup}.

In accordance with the experimental setup of TEM we are looking for solutions
that are slow modulations in the transversal direction $z$ of a periodic
Bloch-type function $\wt \Psi = \sum_{g\in \DLL} \psi_g \ee^{2\pi \ii g\cdot x}$
times the carrier wave $\ee^{\ii\,2\pi k_0\cdot x}$ (multi-beam approach). More
precisely, we seek solutions in the form
\begin{equation}
  \label{eq:Psi.modul}
  \Psi(x)=\Psi(y,z) = \sum_{g\in \DLL} \psi_g(z) \,\ee^{\ii\,2\pi k_0\cdot
    x}\,  \ee^{\ii\,2\pi  g\cdot x}, \quad \text{where }x=(y,z).
\end{equation}
From a physics point of view, this multi-beam ansatz represents the diffraction
of the incoming beam $\psi_0$ in different discrete directions $g$, given by
the dual lattice.  The use of an objective aperture in TEM allows for
restricting the set of transmitted beams forming the image in the
microscope. Bright field and dark field imaging allows us to access the
specific components $\psi_g$ of the multi-beam ansatz.

Using the Fourier expansion of $\calU$ we see that $\Psi$ given in
\eqref{eq:Psi.modul}
solves the Schr\"o\-din\-ger equation \eqref{eq:Mod.Schr} if and only if
the following system of ODEs is satisfied:
\begin{equation}
  \label{eq:Mod.2order}
  \begin{aligned}
    &\pl_z^2 \psi_g(z) + \ii \;\! 4\pi\;\!\rho_g  \pl_z \psi_g(z) +
  4\pi^2\sigma_g \psi_g(z) = -4\pi^2\sum_{h\in \DLL} U_{g-h} \psi_h(z) 
   \quad \text{for } g\in \DLL,\\
  &\text{where } \rho_g:=  (k_0{+}g)\cdot \bfnu \   \text{ and }  \sigma_g :
  = |k_0|^2 - |k_0{+}g|^2 = -|g|^2{-}2 k_0\cdot g.
  \end{aligned}
\end{equation}
Recalling $\bfnu=(0,..,1)^\top$, we see that $\rho_g$ is positive for
$g\approx 0$, while $\sigma_g$ changes sign in balls around $g=0$.  

Next we can use the fact that the variation in $z$ is small such that
$\pl_z^2\psi$ is much smaller than typical values of $\pl_z
\psi_g$. Thus, following the standard practice in TEM (see Remark
\ref{re:JustifyDrop} for the justification), we will neglect the
second derivative and are left with an infinite system of first-order
ordinary differential equations, called \emph{Darwin--Howie-Whelan
  (DHW) equation}, see e.g.\ \cite[Eqn.\,(2.2.1)]{vDyc76IBHE} or
\cite[Eqn.\,(1)]{MNBL19DDEG}.  To simplify notations, we use the
shorthand
$\dot \psi_g(z)=\pl_z \psi_g(z)= \frac{\rmd}{\rmd z} \psi_g(z)$ and
find for the vector $\bfpsi=(\psi_g)_{g\in \DLL}$ the system
\begin{equation}
  \label{eq:Mod.1order}
 \begin{aligned}
    & R \dot\bfpsi = \ii \big( \Sigma + \bbU \big) \bfpsi, \text{ where }\\
  &R=\mathOP{diag}\big(\frac{\rho_g}\pi\big)_{g\in \DLL}, \quad  
  \Sigma =\mathOP{diag}(\sigma_g)_{g\in \DLL}, \quad
  \bbU \bfpsi =\big(\ts\sum_{h\in \DLL} U_{g-h} \psi_h\big)_{g\in \DLL}
\end{aligned}
\end{equation}
Denoting by $\bfdelta:=(\delta_{0,g})_{g\in \DLL}$ (Kronecker symbol) the
incoming beam, the solution $\bfpsi$ of the DHW equations can be written
formally as $\bfpsi(z)= \ee^{\ii R^{-1}(\Sigma +\bbU) z } \bfdelta$.

The following structural assumptions will be fundamental for the
analysis:
\begin{equation}
  \label{eq:Assum.Struct}
\forall\,g\in \DLL:\quad   \rho_g \in \R, \quad \sigma_g \in \R, \quad U_{-g} =
\ol U_g . 
\end{equation}
Hence, the operator $\bbU$ is not only a simple convolution, but it is
additionally Hermitian with respect to the standard complex scalar product.
The latter is crucial for our later analysis. (Sometimes the
Hermitian symmetry of $(U_{g{-}h})_{g,h\in \DLL}$ is broken by adding terms to
model further effects like absorption or radiation. As our approach does
not cover this case, we will not address this point in the present work.)

We emphasize that system \eqref{eq:Mod.1order} has a good structure
because it keeps the symmetries related to self-adjointness of the
Schr\"odinger equation.  However, as is done in the physical literature 
it is often useful, e.g.\ for computational reasons, to write the system as an
explicit first-order equation in the form
\begin{align}
  \label{eq:BasicEquation}
  \dot \bfpsi = \ii \big( 2\pi S {+}\bbW) \bfpsi  \quad& 
 \text{with } S= \mathOP{diag}(s_g)_{g\in \DLL } \text{ and } \text(\bbW
 \bfpsi)_g = \sum_{h\in \DLL } W_{g,h} \,\psi_h,\\ \nonumber
& \text{where } s_g= \sigma_g/(2 \rho_g) \text{ and } W_{g,h}= \pi U_{g-h}/\rho_g. 
\end{align}
The coefficients $s_g$ are called \emph{excitation errors} and play a central
role in TEM. They drive the phase of $\psi_g(z)\in \C$ and can be
interpreted as modulational wave numbers.

The division by the diagonal operator $R=\mathOP{diag}(\rho_g/\pi)$ destroys
two important properties of the operator $\bbU$. 
The scattering operator $\bbW$ is not described by a simple convolution
anymore nor is it Hermitian.
 A serious problem occurs because the factor
$\rho_g = (k_0{+}g)\cdot \bfnu$ may become very small or even exactly $0$. This
happens for $g$ such that $k_0{+}g$ has no component in $z$-direction, i.e.\
the wave travels orthogonal to $\bfnu$. Such waves are not relevant in TEM, and
next we explain below how $g$ is restricted to exclude this case.

\subsection{Restriction to relevant wave vectors}
\label{su:RelevaWaveVect}

The fundamental observation is that the DHW equations for \emph{all}
$g \in \DLL$ is not really what is intended. The equation was derived with the
aim to understand the behavior of $\psi_g$ for $g$ close to $g=0$, because in
high-energy for reasonably thick specimens the diffraction remains small, i.e.\
we should only consider $g$ with $|g|\ll |k_0|$.

Moreover, the assumption that the second derivative
$\pl^2_z \psi_g=\ddot \psi_g$ can be dropped in comparison to the other terms
$\rho_g \dot\psi_g$, $\sigma_g \psi_g$, and $(\bbU \bfpsi)_g$ is only justified
if the excitation error $s_g = \sigma_g/(2\rho_g)$ are small compared to
$1$. Indeed, if $\bbU$ is small with respect to $|k_0|$, which will be one of
our standing assumptions, then ignoring the term with the second derivative in
the left-hand side of
\[
 \frac1{4\pi\rho_g} \ddot \psi_g + \ii \dot\psi_g + 2\pi s_g \psi 
= -\frac{\pi}{\rho_g} (\bbU \bfpsi)_g
\]
leads to the explicit homogeneous solution $\psi_g(z)= \ee^{\ii 2\pi s_g z}$.
The term with the second derivative with respect to $z$ is small relative to
the other terms only if
\begin{equation}
  \label{eq:SlowlyVar}
\begin{aligned}  
\big|\frac1{4\pi\rho_g}  \ddot \psi_g\big| = \frac{\pi s_g^2}{|\rho_g|} 
\; &\ll  \;  | \dot\psi_g|+ |2\pi\;\! s_g \psi_g| = 4\pi|s_g| \\[-0.3em]
&\qquad \qquad   \Longleftrightarrow \ \   |s_g| \ll |\rho_g|
 \ \   \Longleftrightarrow \ \   |\sigma_g| \ll |\rho_g|^2. 
\end{aligned}
\end{equation}
\begin{figure}
\centering
  \begin{tikzpicture}[scale=0.5]
      \fill [color=gray!10]  (-4.2,1.1) rectangle (4.2,-7.2);
      \begin{scope}
      \clip (-4.1,1) rectangle (4.1,-7.1);
       \begin{scope}
         \clip (-4,-3.1) rectangle (4,-7.5) ;
         \fill [color=gray!40] (0,-3.6) circle (3.1);
         \fill [color=gray!10] (0,-2.5) circle (2.9);
       \end{scope}  
       \begin{scope}
         \clip (-4,-2.9) rectangle (4,1) ;
         \fill [color=red!20] (0,-2.4) circle (3.1);
         \fill [color=gray!10] (0,-3.5) circle (2.9);
       \end{scope}  
        \foreach \i in {-18,...,20}
         \foreach \j in {-2,...,20}
          \fill (0.1 +0.4*\i+0.1*\j , -0.4*\j) circle(0.06);
  \end{scope}

   \draw[thick,dashed] (-4.5,-3) -- (5,-3) node[right] {$\rho_g =0$};
   \node [right]at (4.5,-1) {$\rho_g>0$};
   \node [right]at (4.5,-4.5) {$\rho_g<0$};
   \draw[ultra thick, color=blue] (0,-3) circle (3.04);

  \draw[thick,->] (-4.5,0) -- (4.5,0) node[right]{$g_y$} ;
  \draw[thick,->] (0.5,-7.5)-- (0.5,1.5) node[right]{$g_z$}; 
  \fill[color=blue] (0,-3) circle(0.14) node[below]{\colorbox{gray!10}{$-k_0$}};
  \draw[ultra thick, color=blue,->] (0,-3)--(0.5,0) 
      node[pos=0.6,left]{\colorbox{gray!10}{{\larger\color{blue}$k_0$}}} ;
  \end{tikzpicture}
\quad
 \begin{tikzpicture}[scale=0.5]
      \fill [color=gray!10]  (-4.2,1.1) rectangle (4.2,-7.2);
      \begin{scope}
    \fill[color=green!40] (-4.2,-2) rectangle (4.2,1.2);
    \fill [color = red!40] (0.5,0) circle(1.2);
      \clip (-4.1,1) rectangle (4.1,-7.1);
        \foreach \i in {-18,...,20}
         \foreach \j in {-2,...,20}
          \fill (0.1 +0.4*\i+0.1*\j , -0.4*\j) circle(0.06);
  \end{scope}
   \node[color=green!60!black, above] at (-3,1.1) {$\bfG_\gamma$};
    \node[color=red!60!black, above] at (-0.3,1.1) {$\bfG^M$};
   \draw[thick,dashed] (-4.5,-3) -- (5,-3);
   \draw[ultra thick, color=blue] (0,-3) circle (3.04);

  \draw[thick,->] (-4.5,0) -- (4.5,0) node[right]{$g_y$} ;
  \draw[thick,->] (0.5,-7.5)-- (0.5,1.5) node[right]{$g_z$}; 
  \fill[color=blue] (0,-3) circle(0.14) node[below]{\colorbox{gray!10}{$-k_0$}};
  \end{tikzpicture}

  \caption{Ewald sphere $\bbS_\Ew$ (blue) is shown together with the
    points of the dual lattice. Left: The areas around the Ewald sphere show the
    regions where $|\sigma_g| \leq 0.1\,|\rho_g|^2$. Only the upper half with
    $\rho_g>0$ is relevant for the DHW equations. Right: The sets $\bfG_\gamma$
    and $\bfG^M$ lie above the hyperplane $\rho_g=0$ and contain $g=0$. }
\label{fig:Ewald1}
\end{figure}
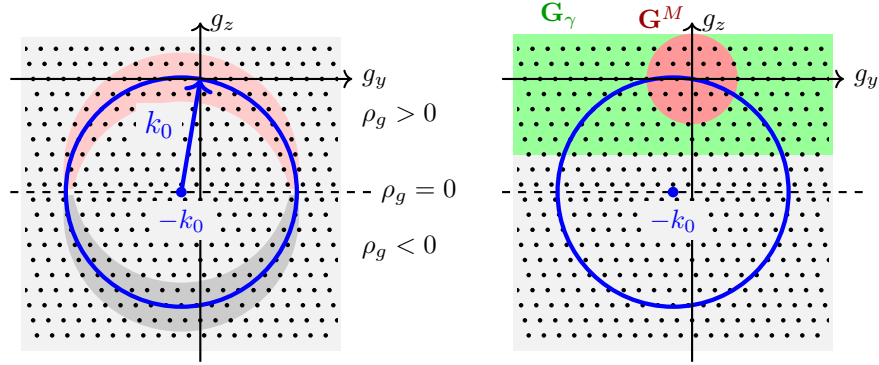

From now on, it will be essential that we restrict the DHW equations 
to a subset $\bfG$ part of the
dual lattice $\DLL$. Two classes of subsets will be used for technical reasons and
exact mathematical estimates, namely
\begin{equation}
  \label{eq:Ggamma.GM}
\bfG_\gamma:= \bigset{g\in \DLL }{ \rho_g \geq \gamma \rho_0} \
\text{and } \  \bfG^M:=B_M(0)\cap \DLL=
\bigset{g\in \DLL}{|g|\leq M}. 
\end{equation}
Throughout we will assume $\gamma \in {]0,1[}$ such that recalling
$\rho_0=k_0\cdot \bfnu>0$ we see that $\bfG_\gamma$ lies above the hyperplane
$\rho_g = (k_0{+}g)\cdot \bfnu=0$ and that $0 \in \bfG_\gamma$ because of
$\gamma \leq 1$. While $\bfG_\gamma$ depends on $k_0$ and contains infinitely
many points, the set $\bfG^M$ is finite and independent of $k_0$. However, we
will always assume $\bfG^M\subset \bfG_\gamma$ for some $\gamma>0$, then
the possible values of $M$ range from $0$ to $\wh m(\gamma,k_0)\approx
(1{-}\gamma)|k_0|$. 

From now on, we will use the shorthand ``\DHW$_\bfG$'' to denote the DHW
equation, where the choice of wave vectors $g$ is restricted to $\bfG$, while
all other $\psi_h$ are ignored, i.e.\ we set $\psi_h\equiv 0$ for $h\not\in\bfG$:
\begin{equation}
  \label{eq:DHW.bfG}
\text{\DHW$_\bfG$:} \qquad   \ii \,\frac{\rho_g}\pi\, \dot \psi_g + \sigma_g \psi_g = -\sum_{h\in \bfG} U_{g-h}
  \psi_h  \quad \text{ for } \ g \in \bfG.
\end{equation}
We will write this equation also in the compact form 
\[
R_\bfG \dot \bfpsi =\ii\!\;\big( \Sigma_\bfG + \bbU_\bfG \big) \;\!\bfpsi
\quad \text{ for } \bfpsi= (\psi_g)_{g\in \bfG}. 
\]
However, whenever possible without creating confusion we will drop the
subscript $\bfG$ and simply write $R$, $\Sigma$, and $\bbU$.    
Throughout we will assume that $0\in \bfG\subset \bfG_\gamma\subset \DLL$ for some
$\gamma \in {]0,1[}$. Our estimates below will show that the difference in
solutions for different sets $\bfG^{(1)}$ and $\bfG^{(2)}$ will be negligible
as long as (i) they both contain a big ball $B_M(0)\cap \DLL$ around $g=0$, (ii)
they are both contained in $\bfG_\gamma$ for some $\gamma\in {]0,1[}$, and (iii) as
long as $z_*$ is not too big, see Corollary \ref{co:ArbiSets}.

\begin{remark}[Justification of dropping  $\pl^2_z \psi_g$]
  \label{re:JustifyDrop}
  In \cite{vDyc76IBHE} the full equation \eqref{eq:Mod.2order}
  including the second-order derivative with respect to $z$ is
  abstractly written as 
\[  
  \frac1{4\pi^2}\ddot \bfpsi + \ii R \dot \bfpsi +(\Sigma {+}\bbU)\bfpsi=0,\quad
  \bfpsi=(\psi_g)_{g\in \bfG}
\]
such that the general solution can be written as the sum
\[
  \bfpsi(z)= \ee^{M_1z} C_1+ \ee^{M_2 z} C_2, \quad \text{where }
  \frac1{4\pi^2}M_j^2 + \ii RM_j +(\Sigma {+}\bbU)I =0
\]
with matrices $M_j$ and  vectors $C_j$. Unfortunately the set $\bfG \subset \DLL$ of
considered wave vectors is not specified. The boundary
conditions are derived in \cite[Sec.\,2.4]{vDyc76IBHE} from the free
equations for $z<0$ and $z>z_*$  (i.e.\ $\bbU=0$) such that
\begin{align*}
&\bfpsi(0)= \bfdelta + \bfpsi_\mafo{reflect},  \quad
\frac1{2\pi^2} \dot\bfpsi(0)+ \ii R \bfpsi(0)= \ii R \bfdelta +
\big( \ii R {+} 2\bbS'\big) \bfpsi_\mafo{reflect}, \\
&\bfpsi(z_*) = \bfpsi_\mafo{transm}, \qquad
\frac1{2\pi^2}\dot\bfpsi(z_*)+ \ii R \bfpsi(z_*)= \big(\ii R{+}2\bbS\big)
\bfpsi_\mafo{transm}. 
\end{align*}
where $\bbS$ and $\bbS'$ are suitable scattering matrices.
It is then shown that $\bfpsi(z)$ differs from
$\bfpsi_\mafo{DHW}(z)=\ee^{\ii R^{-1}(\Sigma {+}\bbU)z} \bfdelta$ only by an
amount that is proportional to $1/|k_0|$, which can be neglected in
most experimentally relevant cases.  
\end{remark}

\section{Mathematical estimates}
\label{se:Estimates}

There are two main reasons that explain why it is possible to replace the
infinite system \eqref{eq:BasicEquation} by a finite-dimensional one. First,
the incoming beam uses only very few modes, usually exactly one.  This means
that the initial condition $\bfpsi(0)$ is strongly localized in the wave-vector
space near $g=0$.  Secondly, as we will explain in the application section (see
Section \ref{se:Applications}), we may assume that the scattering kernel
$ U_{g-h}$ decays exponentially in the distance $|g{-}h|$. Thus, in Section
\ref{su:ExponDecay} we will show that the solution $\bfpsi(z)$ remains
localized on $\DLL$ around the initial beams for all $z\in [0,z_*]$. Thus, we
can show that cutting away modes with $|g|> M$, we obtain an error that decays
like $\ee^{-\lambda M}$ with $\lambda >0$.

The first result concerns the preservation of a specific quadratic form that
can be used as a norm if we restrict the system to a region in $\DLL$ where
$\rho_g>0$. An additional reduction of the number of relevant modes is
discussed in Section \ref{su:Averaging}. It concerns averaging effects that
occur by large excitation errors $s_g=\sigma_g/(2\rho_g)$. The set
\[
  \bbS_\Ew:= \bigset{g \in \R^d }{ |k_0{+}g|^2 = |k_0|^2}
\]
is called the Ewald sphere after \cite{Ewal21BOEG}. For wave vectors
$g \in \DLL$ lying on or near $\bbS_\Ew$ the excitation error $s_g$ is $0$ or
small, respectively. The condition $s_g=0$ means that the Bragg
condition for diffraction holds.

However for $g$ lying far way from $\bbS_\Ew$ we have
$|s_g|\geq s_*\gg 1$, which leads to fast oscillations that make the
amplitudes of these modes small.

\subsection{Conservation of norms}
\label{su:NormPreserved}

We now turn to the analysis of the DHW equations \eqref{eq:DHW.bfG} for a subset
$\bfG$ which may be a system of finite or of infinitely many coupled linear
ODEs. One major impact of restriction to $\bfG\subset \bfG_\gamma$ lies in the
fact that \emph{all} $\rho_g$ are now positive. Thus, we can introduce the norm
\[
\|\psi\|_\bfG := \Big(\sum_{g\in \bfG} \rho_g \, |\psi_g|^2 \Big)^{1/2} = \Big(\big\langle \pi R\bfpsi,\bfpsi\big\rangle \Big)^{1/2} 
\] 
The square $\|\bfpsi\|_\bfG^2$ can be related to the wave
flux in the static Schr\"odinger equation, see Remark \ref{rm:WaveFlux}.  
We define the Hilbert spaces  
\[
\mfH(\bfG):= \bigset{A \in \ell^2(\bfG) }{ \|A\|_\bfG <\infty} \ \text{ with
  scalar product } \big\langle A,B\big\rangle_\bfG := \sum_{g\in \bfG}
\rho_g \,
A_g \ol B_g.
\]

The following classical result states the existence and uniqueness of
solutions for \DHW$_\bfG$ together with the property that the associated
evolution preserves the Hilbert-space norm as well as the energy norm. 

\begin{proposition}[Existence, uniqueness, and conservation of norms]\label{pr:ExUniFlux}
  Assume that $\rho_g$ and $\sigma_g$ are given as in \eqref{eq:Mod.2order} and
  that $\bbU=(U_{g-h})$ satisfies $U_{-g}=\ol U_g$ and
  $|U_g|\leq C_\bbU<\infty$. Then, \DHW$_\bfG$ as given in \eqref{eq:DHW.bfG}
  has for each $\bfpsi(0)\in \mfH(\bfG) $ a unique solution
  $\bfpsi\in \rmC^0(\R;\mfH(\bfG))$. Moreover, the solution satisfies
\begin{equation}
  \label{eq:FluxConserv}
  \|\bfpsi(z)\|_\bfG^2 = \|\bfpsi(0)\|_\bfG^2 \quad \text{and} \quad 
\| H \bfpsi(z) \|_\bfG^2 
 = \| H \bfpsi(0) \|_\bfG^2
\qquad \quad \text{for all } z\in \R. 
\end{equation} 
\end{proposition} 
\begin{proof} 
The result is a direct consequence of the standard theory of the
  generation of strongly continuous, unitary groups $\ee^{\ii z H}$, where
  $H=R^{-1}(\Sigma {+}\bbU)$ is self-adjoint on $\mfH(\bfG)$ equipped with the
  scalar product $\langle \cdot,\cdot\rangle_\bfG$.  
\end{proof}

 The following remark shows that the conservation of the norm $\|\bfpsi\|_\bfG$
is not related to the classical mass conservation in the Schr\"odinger equation
but should be interpreted as a wave-flux conservation, which is only
approximately true in the Schr\"odinger equation, but becomes exact under the
DHW approximation, i.e.\ by ignoring the term involving $\frac{\rmd^2}{\rmd
  z^2} \psi_g$ in \eqref{eq:I.DHW}. 

\begin{remark}[Wave flux in the static Schr\"odinger equation] 
\label{rm:WaveFlux} 
For general solutions $\psi(t,x)$ of the time-dependent Schr\"odinger equation
\ref{eq:I.TDSE} we can introduce the probability density
$\varrho(t,x)=|\psi(t,x)|^2$ and obtain the conservation law
\[
  \frac{\partial \varrho}{\partial t} + \DIV \bfJ = - \frac{2q}{\hbar}
  \Im(V_\rmC)\varrho\quad \text{with } \bfJ = \frac{\hbar}{m} \Im \big(\ol \psi\,
  \nabla \psi \big) \in \R^d,
\]
where $\bfJ$ is called electron-flux vector, see e.g.\
\cite[p.125]{Degr03ICTE}. Because of our assumption \eqref{eq:Assum.Struct} we
have $\Im (V_\rmC)\equiv 0$, such that for solutions $\Psi$ of the static
Schr\"odinger equation the electron flux satisfies $\DIV \bfJ \equiv 0$.

Moreover, by column approximation \eqref{eq:I.ColumnAppr} 
$\Psi(x)=\Psi(y,z)$ is exactly periodic in $y=(y_1,...,y_{d-1})$ and a slowly
varying periodic function in $z$. We denote by
$\calP=\calP'_y{\times}[0,a_0] \subset \R^d$ the periodicity cell of the
crystal, where $a_0$ is the lattice constant. Choosing $z_1,z_2 \in [0,z_*]$
and recalling $\bfnu=(0,...,0,1)^\top$, the divergence theorem gives
\begin{align*}
0 &=\int_{z_1}^{z_2} \int_{(0,z){+}\calP} \DIV \bfJ \dd x \:\dd z =
\int_{z_1}^{z_2} \int_{(0,z)+\pl\calP} \bfJ\cdot \hat n \dd a \,\dd z 
\\
&= \int_{z_1}^{z_2} \Big(\int_{\calP_y\!\ti\! \{z+a_0\}} 
   \hspace*{-1em} \bfJ\cdot \bfnu \dd a  
   - \int_{\calP_y\!\ti\! \{z\}} \hspace*{-1em} 
    \bfJ\cdot \bfnu\dd a \Big) \:\dd z
=  \int_{(0,z_2)+\calP} \hspace*{-1em} \bfJ\cdot \bfnu \dd x 
  - \int_{(0,z_1)+\calP} \hspace*{-1em}  \bfJ\cdot \bfnu \dd x.
\end{align*}
Thus, we conclude that the wave flux $\rmW\rmF(z)$ is independent of $z\in
[0,z_*]$, where 
\[
\rmW\rmF(z) := \int_{(0,z)+\calP} \bfJ \cdot \bfnu \dd x = \frac \hbar m 
 \int_{(0,z)+\calP}
  \Im\big( \ol \Psi  \frac{\pl}{\pl z} \Psi\big) \dd x .
\]

Inserting the Fourier expansion \eqref{eq:I.ColumnAppr} into $
\Im\big( \ol \Psi \frac{\pl}{\pl z} \Psi\big)$ we find that
\begin{align*}
\frac m\hbar \rmW\rmF (z) & = \int_{(0,z)+\calP}
  \Im\big( \ol \Psi(y,\hat z)  \frac{\pl}{\pl \hat z} \Psi(y,\hat z)\big)\, 
    \dd (y,\hat z) \\
& = 
\sum_{g\in \DLL} \Im\Big(\ol \psi_g(z) \,\big(\dot \psi_g(z)+ \ii
\,2\pi(k_0{+}g)\psi_g(z)  \big) \Big)  
\\
& = 2\pi \sum_{g\in \DLL} (k_0{+}g)|\psi_g(z)|^2 \ + \ \sum_{g\in\DLL} \Im\big(\ol
\psi_g(z) \,\dot \psi_g(z)\big). 
\end{align*}
We see that the first sum corresponds to our conserved norm $\| \bfpsi(z)\|^2_\bfG$ if
the contributions of $\psi_g(z)$ for $g\in \DLL{\setminus} \bfG$ are
negligible. The second sum is much smaller than the first sum, because of our
assumption of slowly varying amplitudes, namely $ |\dot \psi_g| \ll |k_0
\psi_g|$, see \eqref{eq:SlowlyVar}. 
\end{remark}

\begin{remark}[Dissipative version of the DHW equations] Often the
  system \eqref{eq:Mod.1order} or \eqref{eq:BasicEquation} are
  modified on a phenomenological level to account for dissipative
  effects like absorption by making $V_\rmC$ complex. Hence, 
  $U_g$ is replaced by $U_g+ \ii
  U'_g$ with a suitable $U'_g$. Under this assumption the above flux
  conservation is no longer true, but most modeling choices (e.g.\ in
  the case that $(U'_{g-h})_{g,h\in \bfG}$ is negative definite) one
  can achieve the estimate $\|\bfpsi(z)\|^2_\bfG \leq \| \bfpsi(0)\|_\bfG^2$
  for $z \geq 0$, i.e.\ the wave flux decays. 
\end{remark}

\subsection{Exponential decay of modes}
\label{su:ExponDecay}

We first show that the solutions can be controlled in an exponentially weighted
norm $\| \cdot\|_\alpha$ with $\alpha \in \R$, where the case $\alpha=0$ would
correspond to the usual norm $\| \cdot\|_\bfG$ in $\mfH(\bfG)$. We define
\[
  \|\bfpsi\|_\alpha^2:= \sum\nolimits_{g\in \bfG} \:\ee^{2\alpha|g|} \rho_g |\psi_g|^2.
\]
Introducing this norm will destroy the Hamiltonian structure of the system. 

\begin{figure}
\centering
\includegraphics[width=1.0\textwidth,trim=0 0 0 0, clip=true]{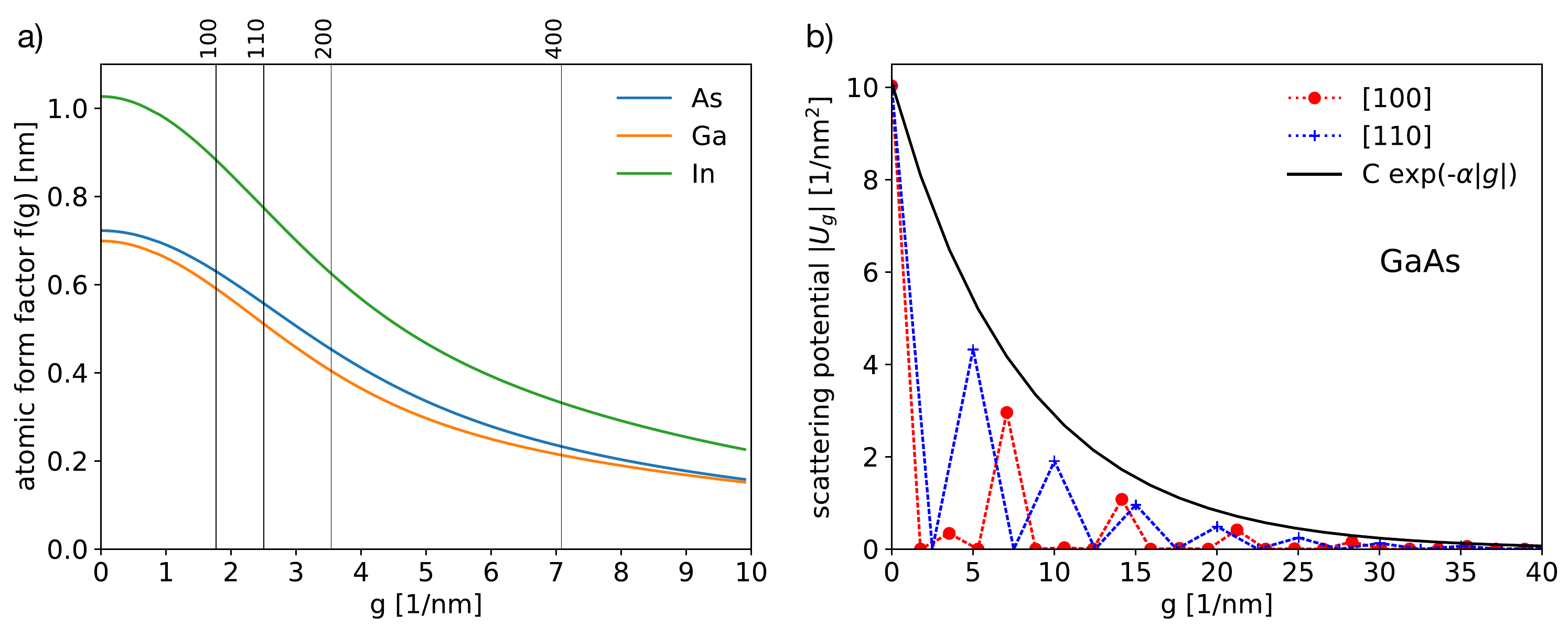}
\caption{Atomic form factors and scattering potential:\  a) atomic form factors
  in dependence on the wave vector $g$ for Ga, As, and In following
  \cite{WeiKo91}.  The vertical lines indicate positions of the lattice planes
  (100), (110), (200) and (400) in GaAs.  \newline b) Fourier coefficients of the
  scattering potential for GaAs along the [100]- (red) and
  [110]-crystallographic directions (blue) as computed by pyTEM
  \cite{Nier19pyTEM} using \eqref{eq:Uscatt_approx}.  The blue and red lines are only
  for guiding of the eye.  An exponential decay (solid black) as assumed in 
  \eqref{eq:Assump.Ug} can be observed with $C=10.1 \, \mathrm{1/nm^2}$ and
  $\alpha=0.125 \ \mathrm{ nm}$.}
\label{fig:uscatt}
\end{figure}

Our main assumption is that the potential operator $\bbU$ acts in
such a way that the Fourier coefficients have exponential decay,
namely
\begin{equation}
  \label{eq:Assump.Ug}
  \exists\, C_\bbU>0,\ \alpha_\bbU>0 \ \forall\, g\in \DLL : \quad
  |U_g|\leq C_\bbU\;\! \ee^{-\alpha_\bbU|g|}. 
\end{equation}
Indeed, the scattering potential can be approximated by 
\begin{equation}
  \label{eq:Uscatt_approx}
U_g \propto \sum\nolimits_\nu \: f_\nu({g}) \ee^{2\pi i g \cdot \bfx_\nu}  \ee^{- M_\nu  |g|^2 },
\end{equation}
where $\bfx_\nu$ denotes the position of the atom $\nu$ in the unit cell,
$f_\nu$ the atomic scattering factors, and 
$M_\nu > 0$ is the Debye-Waller factor, see \cite{WeiKo91} and
\cite{SchoRoetal09},  respectively. 
Thus. assumption \eqref{eq:Assump.Ug} is automatically satisfied.
Figure \ref{fig:uscatt} gives an example for GaAs.

With this assumption we are now able to control the size of the
solutions of \eqref{eq:BasicEquation} in the weighted norm if
$|\alpha| < \alpha _\bbU$. In the following result $\alpha$ may be
positive or negative, but later on we are interested in
$\alpha>0$.

\begin{theorem}[Weighted norms]
\label{th:WeightedEstim}
Consider the \DHW$_\bfG$ as in \eqref{eq:DHW.bfG} with
$\bfG\subset \bfG_\gamma \subset \DLL$ with $\gamma\in {]0,1[}$. Moreover,
assume that $\calU$ satisfies \eqref{eq:Assump.Ug}.  Then, for all
$\alpha \in {]{-}\alpha_\bbU,\alpha_\bbU[}$ and all initial conditions
$\bfpsi^0\in \mfH(\bfG )$ with $\|\bfpsi^0\|_\alpha<\infty$ the unique solution
$\bfpsi$ of \eqref{eq:DHW.bfG} with $\bfpsi(0)=\bfpsi^0$ satisfies the estimate
\begin{equation}
  \label{eq:WeightedEstim}
  \| \bfpsi(z)\|_\alpha \leq  \ee^{\kappa(\alpha) |z|} \|\bfpsi^0\|_\alpha
  \quad \text{for } z \in \R,  
\end{equation}
where the exponential growth rate $ \kappa(\alpha) $ is explicitly given by
\[
  \kappa(\alpha) = \frac{ \pi C_\bbU}{\gamma \rho_0} \:|\alpha|\;\!
  \mfS_1(\alpha_\bbU{-}|\alpha|), \quad \text{where } \mfS_{m}(\beta):=
  \sum_{\varkappa\in \DLL} |\varkappa|^m\ee^{-\beta |\varkappa|}\,.
\]
\end{theorem}
\begin{proof} \STEP{Step 1. Transformation of the equation:} We introduce
  the new variables $B_g= \ee^{\alpha|g|}\psi_g$ such that $\|\bfpsi\|_\alpha =
  \|B\|_\bfG$. In terms of $B$ we can rewrite \DHW$_\bfG$ as 
\begin{equation}
    \label{eq:B.P.alpha}
 \ii R \dot B + \Sigma B = -\bbU^{(\alpha)} B =  -\bbU B + \bbP^{(\alpha)}B \
 \text{ with } \bbP^{(\alpha)}_{g,h} =\big( 1-\ee^{\alpha(|g|-|h|)}\big) U_{g-h}.
\end{equation}
Here we used that $R$ and $\Sigma$ are diagonal operators, and hence commute
with the multiplication of the exponential factor. 
Using the bound \eqref{eq:Assump.Ug} for $U_g$, the coefficients of the
perturbation operator $\bbP^{(\alpha)}$ can be bounded by 
\begin{equation}
  \label{eq:bbP.alpha}
  |\bbP^{(\alpha)}_{g,h}| \leq C_\bbU \big( 1-\ee^{|\alpha|\,|g-h|} \big)
  \ee^{-\alpha_\bbU|g-h|} \leq C_\bbU \,\min\{1,|\alpha|\,|g{-}h|\}\, 
  \ee^{-(\alpha_\bbU{-}|\alpha|)|g-h|}.
\end{equation}

\STEP{Step 2. Operator norm of $R^{-1}\bbP^{(\alpha)}$ in $\mfH(\bfG)$.} 
To control the perturbation $R^{-1}\bbP^{(\alpha)}B$ in terms of
$\|B\|_\bfG$ we employ Lemma \ref{le:OperatorNorm} to obtain
$ \| R^{-1}\bbP^{(\alpha)}B\|_\bfG  \leq C^{(\alpha)}_\bbP \|B\|_\bfG$ with 
\begin{align*}
 C^{(\alpha)}_\bbP:=  \Big( \sup_{g\in \bfG} \sum_{h\in \bfG} 
 \frac{\pi |\bbP_{g,h}^{(\alpha)} |  }{ |\rho_g\rho_h|^{1/2} }  \Big)^{1/2}  
     \Big( \sup_{h\in \bfG} \sum_{g\in \bfG} \frac{\pi |\bbP_{g,h}^{(\alpha)} |  }{
  |\rho_g\rho_h|^{1/2} }  \Big)^{1/2}.
\end{align*}
Because of $\rho_g,\rho_h \in \bfG\subset \bfG_\gamma$, and
\eqref{eq:bbP.alpha} this yields   
\begin{align*}
C^{(\alpha)}_\bbP & \leq  \pi C_\bbU \: \sup_{g\in \bfG} \sum_{h\in \bfG}
\frac{|\alpha|\, |g{-}h|}{\gamma \rho_0}  \: \ee^{-(\alpha_\bbU{-}|\alpha|)|g-h|} 
\ \leq \  \frac{\pi C_\bbU} {\gamma \rho_0}\:|\alpha|\!\; 
 \mfS_1(\alpha_\bbU{-} \alpha)\ = \ \kappa(\alpha). 
\end{align*}

\STEP{Step 3. Gronwall estimate.}
We now apply the variation-of-constants formula (Duhamel's principle) to the
solution $B$ for \eqref{eq:B.P.alpha} in the $\mfH(\bfG)$, where
$H=R^{-1}(\Sigma{-}\bbU)$ is the generator of the norm-preserving $C_0$ group
$\ee^{\ii z H}$:
\begin{align*}
  \|B(z)\|_\bfG &\leq  \| \ee^{iH z} B(0)\|_\bfG +\int_0^z \|
  \ee^{i(z-\zeta)H } \|_\bfG \| R^{-1}\bbP^{(\alpha)} B(\zeta) \|_\bfG \dd
  \zeta 
\\
&\leq
  \| \bfpsi(0)\|_\alpha + \int_0^z  \kappa(\alpha)\, \|B(\zeta)\|_\bfG \dd \zeta .
\end{align*}
Now, Gronwall's estimate gives $\|\bfpsi(z)\|_\alpha= \|B(z)\|_\bfG \leq
\ee^{\kappa(\alpha) z} \|\bfpsi(0)\|_\alpha$, and the proof is completed. 
\end{proof}

Step 2 of the above proof relies on the following elementary lemma, which will
be used again to calculate the norm of convolution-type operators involving
$\bbU$.

\begin{lemma}[Operator norm]\label{le:OperatorNorm} Consider
  $\bfG^{(1)},\bfG^{(2)}\subset \bfG_\gamma$ with $\gamma>0$ and $\bbB:\mfH(\bfG^{(1)})
  \to \mfH(\bfG^{(2)})$ with $(\bbB A )_g =
  \sum_{h\in \bfG^{(1)}} B_{gh}A_h$. Setting $\wt B_{gh}= \sqrt{\rho_g/\rho_h} \,
  B_{gh}$ gives
\begin{equation}
\label{eq:row-col-norm}
\| \bbB A\|_{\bfG^{(2)}} \leq C_\bbB \|A\|_{\bfG^{(1)}} ,\ \ C_\bbB= 
\Big(\! \sup_{h\in \bfG^{(1)}}\! \sum_{g\in \bfG^{(2)}}\!\! |\wt B_{gh}|\Big)^{1/2}  
  \Big( \!\sup_{g\in \bfG^{(2)}} \!\sum_{h\in \bfG^{(1)}}\!\! |\wt B_{gh}|\Big)^{1/2}\!,
\end{equation}
which is the square root of the product of the row-sum and column-sum
norm.
\end{lemma} 
\begin{proof} With $r_g =\rho_g^{1/2}$ the desired result is obtained as follows:
\begin{align*}
\| \bbB A\|_{\bfG^{(2)}}^2&= \!\! \sum_{g\in \bfG^{(2)}}\!\!  r_g^2 
  \Big(\!\sum_{h\in \bfG^{(1)}}\!\!\!   B_{gh} A_h\Big)
\ol{(\bbB A)}_g  \leq  \!\! 
 \sum_{g\in \bfG^{(2)}} \sum_{h\in \bfG^{(1)}} \!\!\! r_h |\wt B_{gh}|^{1/2} 
   |A_h| r_g |\wt B_{gh}|^{1/2} |(\bbB A)_g| \\
&\leq_{\mafo{CaSch}} \Big( \sum_{g\in \bfG^{(2)}} \sum_{h\in \bfG^{(1)}}  \!\! 
  |\wt B_{gh}|\, r_h^2|A_h|^2 \Big)^{1/2}  \Big( \sum_{g\in \bfG^{(2)}} 
   \sum_{h\in \bfG^{(1)}}  \!\! |\wt B_{gh}|\, r_g^2|(\bbB A)_g|^2 \Big)^{1/2} \\
& \leq \ \Big(\sup_{h\in \bfG^{(1)}} \big(\sum_{g\in \bfG^{(2)}}  \!\! |\wt B_{gh} | 
   \big)\Big)^{1/2}  \|A\|_{\bfG^{(1)}}  \ \Big(\sup_{g\in \bfG^{(2)}} 
 \big(\sum_{h\in \bfG^{(1)}} \!\! |\wt B_{gh}| \big)\Big)^{1/2} \,\|\bbB A\|_{\bfG^{(2)}}
  \\ 
& \ = \ C_\bbB \| A\|_{\bfG^{(1)}}\|\bbB A\|_{\bfG^{(2)}}.  
\end{align*}
Thus, Lemma \ref{le:OperatorNorm} is established. 
\end{proof}

The importance of Theorem \ref{th:WeightedEstim} is that the growth rate
$\kappa(\alpha)$ is completely independent of the domain $\bfG$ as long as
$\bfG$ is contained in $\bfG_\gamma$. Thus, we will have the option to compare
solutions obtained for different wave-vector sets $\bfG$. 

As a first consequence we obtain that for all $z\in [0,z_*]$ the solutions
$\bfpsi(z)=(\psi_g(z))_{g\in \bfG}$ decay with $|g|\to \infty$. Indeed,
recalling that the initial condition is given by the incoming wave encoded in
the $\bfdelta=(\delta_{0,g})_{g\in \bfG}$ (Kronecker symbol) we have 
\begin{equation}
  \label{eq:InitialNorm}
  \|\bfpsi(0)\|_\alpha = \|\bfpsi(0)\|_\bfG = \|\bfdelta\|_\bfG =
  \sqrt{\rho_0} = \sqrt{k_0\cdot\bfnu} \approx \sqrt{|k_0|},
\end{equation}
which is independent of the exponential weighting by $\alpha$.  
With this and $\alpha\in {]0,\alpha_\bbU[}$ we obtain
\[
|\psi_g(z)| \ \leq \ \frac{\ee^{-\alpha|g|}}{\sqrt{\rho_g}}
\|\bfpsi(z)\|_\alpha  \ \leq \ \ee^{\kappa(\alpha) |z| -\alpha|g|}
\sqrt{\frac{\rho_0}{\rho_g}}.   
\]
Thus, the exponential factor $\ee^{\kappa(\alpha) |z| -\alpha|g|} $ shows that
the solution $\bfpsi(z)$ can only have a nontrivial 
effect at $g\neq 0$ if $|z| > \alpha/\kappa(\alpha) \:|g|$.  We may
consider the quotient $\alpha/\kappa(\alpha)$ as a collective scattering length
that describes how fast a beam has to travel through the specimen 
to generate nontrivial amplitudes at neighboring wave vectors $g$. 
In contrast, the extinction length $\xi_g$ is defined for
each individual $g \in \DLL$ (see \cite[p.309]{Degr03ICTE}):
\[
\frac{\alpha}{\kappa(\alpha)} = \frac{\gamma\, k_0\cdot\bfnu}{\pi C_\bbU \mfS_1
 (\alpha_\bbU{-}\alpha)} \quad \text{versus the extinction length }
\xi_g:= \frac{ |\rho_g|}{|U_g|}.    
\]

Hence, for doing a reasonable TEM experiment, one wants $\kappa(\alpha)z_*$ to
be big enough to see some effect of scattering. However, it should not be too
big such that the radius of possibly activated wave vectors with $|\psi_g|\geq \eps$
is not too small, namely those with $|g| \leq \frac1\alpha\big(\kappa(\alpha)z_*
+ \log(1/\eps)\big)$. In addition we define the excitation length to be
$\ell_\mafo{excit}(s_g)=1/ |s_g|$, which describes the period of the phase
oscillation of $\psi_g(t)$.

\subsection{Error estimates for finite-dimensional approximations}
\label{su:CutOffError}

We now compare the DHW equations on different sets $\bfG^{(1)}$ and
$\bfG^{(2)}$, both contained in $\bfG_\gamma \subset \DLL$.  We denote by
$\bfpsi^{(j)}$ the solution of \DHW$_{\bfG^{(j)}}$ with the initial condition
$\bfpsi^{(j)}(0)=\bfdelta$.  

Assuming $\bfG^{(1)} \subset \bfG^{(2)} $ we can decompose $\bfpsi^{(2)}$ into
two pieces, namely 
\[ 
  \bfpsi^{(2)} = (B,C)\quad \text{with }B=\bfpsi^{(2)}|_{\bfG^{(1)}}
  :=(\psi_g)_{g\in \bfG^{(1)}} \text{ and }
  C=\bfpsi^{(2)}|_{\bfG^{(2)}\setminus \bfG^{(1)}} 
  .
\]
We may rewrite the \DHW$_{\bfG^{(2)}}$ in block structure via
\begin{subequations}
  \label{eq:DHW.block}
  \begin{align}
      \label{eq:DHW.block.B}
   R_{(1)} \dot B \ \ &=\ii\big(\ \ \Sigma_{(1)} B \ \ + \bbU_{BB} B + \bbU_{BC} C\big),
  &&B(0) = (\delta_{0,g})_{g\in \bfG^{(1)}}        
    \\
      \label{eq:DHW.block.C}
  R_{(2)\setminus(1)} \dot C &=\ii\big(\Sigma_{(2)\setminus(1)} B 
          +\bbU_{CB} B + \bbU_{CC} C\big),
  && C(0)=0  .
  \end{align}
\end{subequations}
Here we used that the initial conditions $\bfpsi(0)$ is localized in the
incoming beam such that $\psi_g(0)=0$ for $g \in \bfG^{(2)}\setminus
\bfG^{(1)}$. Moreover, the \DHW$_{\bfG^{(1)}}$ is given by 
\eqref{eq:DHW.block.B} if we omit the coupling term
`` $+\bbU_{BC}C$ '':
\begin{equation}
  \label{eq:DHW.wtB}
   R_{(1)}M \dot\bfpsi^{(1)}= \ii\,\big( \Sigma_{(1)} + \bbU_{BB} \big)
  \bfpsi^{(1)} , \quad 
   \bfpsi^{(1)}(0) = (\delta_{0,g})_{g\in \bfG^{M}}   . 
\end{equation}

The following result provides a first estimate between the solution
$\bfpsi^{(2)}=(B,C)$ on the larger wave-vector set $\bfG^{(2)}$ and
$\bfpsi^{(1)}$ on the smaller set $\bfG^{(1)}$ by exploiting the exponential
decay estimates established in Theorem \ref{th:WeightedEstim}.  In this first
case, we consider only the model sets $\bfG^{(1)}=\bfG^{M} $ and
$\bfG^{(2)}= \bfG_\gamma$ with $\bfG^M\subset \bfG_\gamma$, see
\eqref{eq:Ggamma.GM}.

\begin{theorem}[Control of approximation errors] Assume \label{th:ApprError} 
that the assumptions \eqref{eq:Assum.Struct} and \eqref{eq:Assump.Ug}
hold and that $k_0$, $M$ and $\gamma \in {]0,1[}$ are such that $\bfG^M\subset
\bfG_\gamma$.  Then, for $\alpha \in {]0,\alpha_\bbU[}$
the solutions $\bfpsi^\gamma$ and $\bfpsi^M$ of \DHW$_{\bfG_\gamma}$  and
\DHW$_{\bfG^M}$ with initial condition $\bfdelta$ satisfy the estimates

\begin{subequations}
  \label{eq:ApproxErr}
  \begin{align}
     \label{eq:ApproxErr.B}
    \big\| \bfpsi^M(z)-\bfpsi^\gamma(z)|_{\bfG^M} \big\|_{\bfG^M}
    &\leq  \frac{ \mfS_0(\alpha_\bbU) -1 }{ \alpha \,\mfS_1(\alpha_\bbU{-}\alpha )}
      \:\ee^{\kappa(\alpha)|z| -\alpha M} \:\|\bfdelta\|_0 \ \ \text{ and } 
    \\[0.4em]
    \big\|\bfpsi^\gamma(z)|_{\bfG_\gamma \setminus \bfG^M} 
     \big\|_{\bfG_\gamma \setminus \bfG^M} \ \ 
    &  \label{eq:ApproxErr.C}
    \leq \ee^{\kappa(\alpha)|z|- \alpha M} \|\bfdelta\|_0 \quad \text{for
      all } z \in \R.                     
  \end{align}
\end{subequations}
where as before $ \mfS_m(\beta) := 
\sum_{\varkappa\in \DLL}|\varkappa|^m \,\ee^{-\beta |\varkappa|}$.
\end{theorem} 
\begin{proof} We denote by $\bfG_\rmO:= \bfG_\gamma\setminus \bfG^M$
  the set of outer wave vectors. 

\STEP{Step 1: Estimate of $C$.} The solution $\bfpsi^\gamma=(B,C)$ satisfies all
assumptions of Theorem \ref{th:WeightedEstim}. Hence, we can rely on the
exponential estimate and obtain 
\begin{align*}
\|C(z)\|_{\bfG_\rmO}^2 \ & 
= \ \sum_{h\in \bfG_\rmO} \rho_h\, |\psi_h(z)|^2 
\ \leq \ \ee^{-2\alpha M} \sum_{g\in \bfG_\gamma} \rho_g\,
\ee^{2\alpha|g|} \,|\psi_g(z)|^2 
\\
& = \ \ee^{-2\alpha M}\|\bfpsi^M(z)\|_\alpha^2 \ \leq \  \ee^{2\kappa(\alpha) |z|
    -2\alpha M} \| \bfpsi^M(0)\|_\alpha^2\ = \  \ee^{2\kappa(\alpha) |z|
    -2\alpha M} \| \bfdelta\|_0^2,
\end{align*}
which is the desired result \eqref{eq:ApproxErr.C}. 

\STEP{Step 2. Estimate between $B$ and $\bfpsi^M$.}
For comparing $B$ and $\bfpsi^M$ we define $A=B-\bfpsi^M$ and see that $A$
satisfies
\begin{equation}
  \label{eq:ODE.Differ.a}
    R_M \dot A (z)=\ii \big( \Sigma_M A(z)+ \bbU_{BB} A(z) + \bbU_{BC}
    C(z)\big),  \qquad   A(0)=0,
\end{equation}
where now the initial condition is $0$. Using the unitary group $\ee^{\ii z
  H_M}$ on $\mfH(\bfG^M)$ defined via $H_M=R_M^{-1}(\Sigma_M{+}\bbU_{BB})$,
the solution is given in terms of Duhamel's principle via $
A(z)= \int_0^z \ee^{\ii(z{-}\zeta)H_M } R_M^{-1} \bbU_{BC} C(\zeta) \dd
\zeta$. Taking the norm in $\mfH(\bfG^M)$ we arrive at 
\[
\|A(z)\|_{\bfG^M} 
 \leq \! \int_0^z \!\!\| R_M^{-1} \bbU_{BC} C(\zeta)\|_{\bfG^M} \dd \zeta 
 \leq \| R_M^{-1}\bbU_{BC}\|_{\mfH(\bfG_\rmO){\to}\mfH(\bfG^M)}  
       \! \int_0^z \!\|C(\zeta)\|_{\bfG_\rmO} \dd \zeta. 
\]
Using Lemma \ref{le:OperatorNorm}, the operator norm
$N_\mafo{cpl}=\|R_M^{-1}\bbU_{BC}\|$ can be estimated by 
\[
  N_\mafo{cpl} \leq \Big( \sup_{h\in \bfG^M} \!\sum_{g\in \bfG_\rmO}\!\!\!
  \tfrac{\pi|U_{g-h}|}{\sqrt{\rho_g\rho_h} }\Big)^{1/2} \Big( \sup_{g\in
    \bfG_\rmO} \!\sum_{h\in \bfG^M}\!\!\!
  \tfrac{\pi|U_{g-h}|}{\sqrt{\rho_g\rho_h} }\Big)^{1/2} \! \leq
  \tfrac{\pi}{\gamma \rho_0} \sup_{g\in \DLL} \!\sum_{h\in
    \DLL\setminus\{g\}}\!\!\!\! C_\bbU \ee^{-\alpha_\bbU|g{-}h|},
\]  
where we used $g\in \bfG_\gamma$  and \eqref{eq:Assump.Ug}. We also 
increased the sets $\bfG^M$ and $\bfG_\rmO$ but kept the information
that they are disjoint by summing only over $h$ different from $g$.  Thus, we
find $N_\mafo{cpl}\leq \pi C_\bbU\big( \mfS_0(\alpha_\bbU)-1\big)/(\gamma \rho_0)$. 

Inserting this into the bound for $\|A(z)\|_{\bfG^M} $ and integrating the
bound \eqref{eq:ApproxErr.C} 
for $C(\zeta)$ we see cancellations in the factor $
N_\mafo{cpl}/ \kappa(\alpha)$, and the result \eqref{eq:ApproxErr.B} follows.
\end{proof}

The next result is dedicated to the case of two general sets $\bfG^{(1)}$ and
$\bfG^{(2)}$ both of which satisfy
$\bfG^M \subset \bfG^{(j)} \subset \bfG_\gamma$ for $j=1,2$. Denoting by
$\bfpsi^{(j)}:[0,z_*]\to \mfH(\bfG^{(j)})$ the solutions of
\DHW$_{\bfG^{(j)}}$, we will see that their restrictions to $\bfG^M$ will be
exponentially close with a factor $\ee^{-\alpha M}$. This explains why the
exact choice of the subset $\bfG$ of the wave vectors is not relevant for
$z\in [0,z_*]$ as long
as it contains a sufficiently large subset $\bfG^M$, i.e.\ $M$ is much larger
than $\kappa(\alpha) z_*/\alpha$.

\begin{corollary}[Arbitrary sets $\bfG^{(j)}$ of wave vectors]\label{co:ArbiSets}
Consider $k_0$, $\gamma$, and $M$ as in Theorem \ref{th:ApprError} and consider two
subsets $\bfG^{(j)} \subset \DLL$ satisfying $\bfG^M \subset \bfG^{(j)} \subset
\bfG_\gamma$ for $j=1,2$. Then, the solutions $\bfpsi^{(j)}$ of
\DHW$_{\bfG^{(j)}}$ with initial condition $\bfpsi^{(j)}(0)=\bfdelta$ satisfy the estimate 
\[
\big\| \bfpsi^{(1)}(z)|_{\bfG^M} -  \bfpsi^{(2)}(z)|_{\bfG^M} \big\|_{\bfG^M}
\leq \frac{ 2\:\big(\mfS_0(\alpha_\bbU)-1\big) } 
 { \alpha \,\mfS_1(\alpha_\bbU{-}\alpha )} \: 
 \ee^{\kappa(\alpha)|z| - \alpha M}\,\|\bfdelta\|_0  \quad
    \text{for all } z\in \R\,. 
\]
\end{corollary} 
\begin{proof} The proof follows simply by observing that Theorem
  \ref{th:ApprError} can easily be generalized by replacing the bigger set
  $\bfG_\gamma$ by any set $\bfG$ between $\bfG^M$ and $\bfG_\gamma$. Hence, we can
  compare the two solutions $\bfpsi^{(j)}$ on $\bfG^M$ with the solution $\bfpsi^M
  $ of \DHW$_{\bfG^M}$. Now the result follows by 
\[
\| \bfpsi^{(1)}|_{\bfG^M} -  \bfpsi^{(2)}|_{\bfG^M} \|_{\bfG^M}
\leq \| \bfpsi^{(1)}|_{\bfG^M} - \bfpsi^M\|_{\bfG^M}
  + \| \bfpsi^M  -  \bfpsi^{(2)}|_{\bfG^M} \|_{\bfG^M}
\]
and applying \eqref{eq:ApproxErr.B} with
$\|\bfpsi(0)\|_\alpha=\|\bfdelta\|_0 =\rho_0^{1/2}$. 
\end{proof}

From now on we will always choose $\alpha=\alpha_\bbU/2$, which is the
intermediate value that makes all sums $\mfS_m\big(\alpha_\bbU/2\big)$ finite. 
Thus, the critical exponential error term takes the form 
\[
\ee^{\kappa(\alpha_\bbU/2)z- M\alpha_\bbU/2} \quad \text{for } z\in [0,z_*].
\]

From practical purposes there is no reason of doing a calculation in a set
$\bfG$ bigger than $\bfG^M$, since increasing the number of ODEs without a gain
in accuracy is useless. Moreover, it is desirable to reduce $M$
as much as possible as the number of ODEs in \DHW$_{\bfG^M}$ with $M=\mu|k_0|$
is proportional to $M^d$. However, in a true experiment we want to see the
effect of scattering such that $\kappa(\alpha)z_*$ needs to be big enough. 
The way to make this work is to choose $M$ proportional to a small power of
$|k_0|$:
\[
M \sim |k_0|^\eta \quad \text{with } \eta \in {]0,1[}.
\]  
For instance the first few Laue zones (see below) can be obtained
by $\eta=1/2$. 
  
While in a ball $\bfG^M$ of radius $M$ the number of wave vectors  scales with
$M^d$, there are further physical reasons that many of these wave vectors are
not relevant, as they cannot be activated because of energetic criteria as
discussed now.

\subsection{Averaging via conservation of the energy norm}
\label{su:Averaging} 

The relevance of the Ewald sphere lies in the fact that on $\bbS_\Ew$ the
excitation error $s_g= \sigma_g/(2\rho_g)$ equals $0$. This means that beams
propagating with wave vectors $g \in \bbS_\Ew$ have much lower energy, because
they have only little phase oscillations. Beams with wave vectors that
are not close to the Ewald sphere will necessarily have much smaller amplitudes,
because they have much larger phase oscillations than beams with wave vectors near the
Ewald sphere. Mathematically this can be manifested by conservation of suitable
energies. Another way of obtaining this result would be by performing a
temporal averaging for the functions $\psi_g$ with large $|s_g|$. 

We return to the DHW equations on $\bfG=\bfG^M \subset \bfG_\gamma$. The
linear finite-dimensional Hamiltonian system 
\begin{equation}
  \label{eq:FullSyst}
   R \dot \bfpsi = \ii \big( \Sigma {+}\bbU \big)\bfpsi, \quad \bfpsi(0)= \bfdelta
  \in \mfH(\bfG^M),
\end{equation}
can be rewritten via the 
transformation $\wt R= R^{1/2} = \mathOP{diag}\big((\rho_g/\pi)^{1/2}\big)_{g\in
  \bfG} $. Setting $A= \wt R \bfpsi$, system  \eqref{eq:FullSyst} takes the standard
Hamiltonian form 
\begin{equation}
  \label{eq:StandardForm}
  \dot A = \ii \,\bbH  A \quad \text{with } \bbH=  \wt R^{-1} \, 
  \big( \Sigma {+} \bbU\big)\, \wt R^{-1},
\end{equation}
where $\bbH$ is now a Hermitian operator on $\ell^2(\bfG^M)$, now using the
standard scalar product.  This provides the explicit solution
$A(z)=\ee^{\ii z \bbH }A(0)$ via the unitary group
$z\mapsto \ee^{\ii z \bbH }$. An easy consequence is the invariance of the
hierarchy of norms:
\[
\forall\, k\in \N_0\ \forall \, z\in \R: \quad 
 \big\langle \bbH^k A(z), A(z)\big\rangle  =  
\big\langle \bbH^k A(0), A(0)\big\rangle \,.
\]
For $k=0$ this is the simple wave-flux conservation established in  Proposition
\ref{pr:ExUniFlux}.  The result for $k=1$ is not useful, because the
operator $\bbH$ is indefinite, since $\bbU$ is bounded and $\Sigma$ has many 
positive (associated with $g$ inside the Ewald sphere) and many negative
eigenvalues (associated with $g$ outside the Ewald sphere). 

Hence, we concentrate on the case $k=2$, where 
\[
0 \leq \bbH^2 = \wt\Sigma^2 +\wt\bbU \wt\Sigma + \wt\Sigma \wt\bbU  + \wt \bbU^2,
\quad \text{with } \wt\Sigma: =  \wt R^{-1}\Sigma \wt R^{-1} =
R^{-1}\Sigma \text{ and } \wt\bbU:=  \wt R^{-1}\bbU \wt R^{-1}.
\]

The following, rather trivial result highlights that $\bbH^2$ has suitable
definiteness properties that will then be useful for estimating the solutions
of the DHW equations.

\begin{lemma}[Energy estimate]\label{le:bbH.vs.Sigma} Let $\bbH=\wt\Sigma+
  \wt\bbU$ where $\wt\Sigma$ and $\bbH$ are Hermitian, then we have the
  estimate
\begin{align}
\label{eq:Estim.bbH2}
\| \bbH A\|^2 &=\big\langle \bbH^2 A, A\big\rangle \geq  
  \frac12 \|\wt\Sigma A\|^2 -  \| \wt\bbU A\|^2 
    \quad \text{for all } A\in \ell^2(\bfG^M). 
\end{align}
\end{lemma}
\begin{proof} We expand $\bbH^2$ in a suitable way:
\begin{align*} 
\bbH^2&= (\wt\Sigma + \wt\bbU)^2 = \frac12\wt\Sigma{}^2 +
\big(\frac12\wt\Sigma{}^2 + \wt\Sigma \wt\bbU + \wt\bbU \wt\Sigma +
2\wt\bbU{}^2\big) - \wt\bbU{}^2\\
& = \frac12\wt\Sigma{}^2 +
\Big(\frac1{\sqrt2}\wt\Sigma + \sqrt2 \wt\bbU\Big)^2 - \wt\bbU{}^2 \ \geq
\  \frac12\wt\Sigma{}^2- \wt\bbU{}^2.
\end{align*}
This is the desired result.  
\end{proof} 

It is instructive to transform estimate \eqref{eq:Estim.bbH2} back to the
original variable $\bfpsi$ and the operator $H=R^{-1}(\Sigma{+}\bbU)$, which yields
\begin{equation}
  \label{eq:H.Sigma.bfpsi}
  \| R^{-1}\Sigma \bfpsi\|_{\bfG^M}^2 \ \leq \ 2\, \|H\bfpsi\|^2_{\bfG^M} +
   2\, \| R^{-1} \bbU \bfpsi\|_{\bfG^M}^2. 
\end{equation}
Since along solutions $z\mapsto \bfpsi(z) \in \mfH(\bfG^M)$ of \DHW$_{\bfG^M}$
the energy $ \|H\bfpsi(z)\|^2_{\bfG^M}$ is constant, 
see \eqref{eq:FluxConserv},
we can use this for
bounding $\|R^{-1}\Sigma \bfpsi(z)\|_{\bfG^M}^2$. 

\begin{proposition}[Energy bound for solutions]\label{pr:SolEnerBou}
Consider $k_0$, $\gamma$, and $M$  such that $\bfG^M\subset \bfG_\gamma$. Let
$\bfpsi$ be the solution of \DHW$_{\bfG^M}$ with initial condition 
$\bfpsi(0)=\bfdelta$. Then, $\bfpsi$ satisfies the estimate 
\[
 \| R^{-1}\Sigma \bfpsi(z)\|_{\bfG^M} \leq   \frac{2 \pi\;\! C_\bbU
   \mfS_0(\alpha_\bbU)}{ \gamma \rho_0} \| \bfdelta\|_{\bfG^M} 
  \quad  \text{for all } z\in \R. 
\]
\end{proposition}
\begin{proof} 
Lemma \ref{le:OperatorNorm} yields $\|R^{-1}\bbU
\bfpsi\|_{\bfG^M} \leq \NNeins{} \|\bfpsi\|_{\bfG^M} $ with $\NNeins{}= \pi\;\! C_\bbU
\mfS_0(\alpha_\bbU) /(\gamma\rho_0)$. Exploiting \eqref{eq:H.Sigma.bfpsi} and
the conservation of $ 
  \|H\bfpsi(z)\|^2_{\bfG^M}$ and $\|\bfpsi(z)\|_{\bfG^M}^2$ we find
\begin{align*}
 \| R^{-1}\Sigma \bfpsi(z)\|_{\bfG^M}^2 & \leq 
   2\|H\bfpsi(z)\|^2_{\bfG^M}  {+} 2\NNeins2 \|\bfpsi(z)\|_{\bfG^M}^2\\
& =  2\|H\bfpsi(0)\|^2_{\bfG^M} {+} 2\NNeins2 \|\bfpsi(0)\|_{\bfG^M}^2 
= 2\|R^{-1}\bbU \bfdelta\|^2_{\bfG^M} {+} 2\NNeins2 \rho_0  \leq  4\NNeins2 \rho_0,
\end{align*}
where we used $\bfpsi(0)=\bfdelta$ and $\sigma_0=0$. This shows the desired assertion. 
\end{proof}

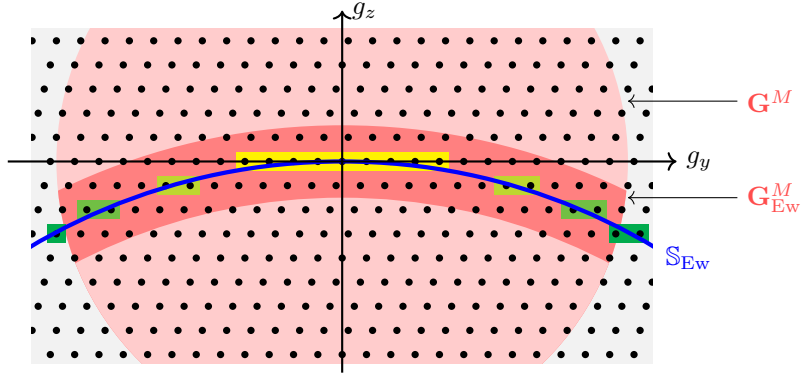
\begin{figure}
\centering
  \begin{tikzpicture}[scale=0.8]
    \begin{scope}
      \clip (-5.11,-3.35) rectangle (5.11,2.2);
      \fill [color=gray!10]  (-5.5,-3.5) rectangle (5.5,2.5);
      \fill [color=red!20] (0,0) circle (4.7);
      \begin{scope}
        \clip (0,0) circle (4.7);
        \fill [color=red!50] (0,-10) circle (10.6);
        \fill [color=red!20] (0,-10) circle (9.4);
      \end{scope}  
       \fill [color=yellow] (-1.75,-0.15) rectangle (1.75,0.15); 
       \fill [color=yellow!70!green] (-3.05,-0.55) rectangle (-2.35,-0.25);
       \fill [color=yellow!70!green] (2.5,-0.55) rectangle (3.25,-0.25);
       \fill [color=yellow!40!green] (-4.35,-0.95) rectangle (-3.65,-0.65);
       \fill [color=yellow!40!green] (3.6,-0.95) rectangle (4.35,-0.65); 
       \fill [color=yellow!10!green] (-4.85,-1.35) rectangle (-4.55,-1.05);
       \fill [color=yellow!10!green] (4.4,-1.35) rectangle (5.05,-1.05); 
       \foreach \i in {-18,...,20}
         \foreach \j in {-9,...,6}
          \fill (0.4*\i+0.1*\j , 0.4*\j) circle(0.06);
       \draw[ultra thick, color=blue] (0,-10) circle (10);
    \end{scope}
  \node[color=blue] at (5.7,-1.6){$\bbS_\Ew$};
  \draw[->] (6.5,-0.6) node[right, color= red!70]{$\bfG^M_\Ew$} -- (4.7,-0.6);
  \draw[->] (6.5,1) node[right, color= red!70]{$\bfG^M$} -- (4.7,1);  
  \draw[thick,->] (-5.5,0) -- (5.5,0) node[right]{$g_y$} ;
  \draw[thick,->] (0,-3.5) -- (0,2.5) node[right]{$g_z$}; 
  \end{tikzpicture}
  \caption{Ewald sphere $\bbS_\Ew$ (blue), dual lattice $\DLL$ (black dots),
    and the decomposition of $\bfG^M$ (light red ball) into $\bfG^M_\Ew$ (red
    annular arc, for $\wt s_*=3\pi$) and $\bfG^M_\far$. The boxes indicate the
    Laue zones: lowest order (yellow) to third order (green). }
\label{fig:Ewald}
\end{figure}

Using the energy bound, we can split the set $\bfG^M$ according to the size of
the excitation errors $s_g=\sigma_g/(2\rho_g)$ using a cut-off value $\wt s_*$
to be chosen later:
\begin{align*}
\bfG^M= \bfG^M_\Ew \,\dot\cup\, \bfG^M_\far \quad \text{with }& 
\bfG^M_\Ew:=\bigset{g\in  \bfG^M}{ |\sigma_g|/(2\rho_g) < \wt s_*} \\
 \text{ and }&
\bfG^M_\far\,:=\bigset{g\in  \bfG^M}{ |\sigma_g|/(2\rho_g) \geq \wt s_*}\,.
\end{align*}
Of course, we always have $g=0 \in \bfG^M_\Ew$, as $\sigma_0=0$ and $\wt s_*>0$.

Using the energy bound from Proposition \ref{pr:SolEnerBou} and $R^{-1}\Sigma =
\mathOP{diag}(\pi \sigma_g/\rho_g)$ we immediately see
that solutions $\bfpsi$ of \DHW$_{\bfG^M}$ satisfy
\begin{equation}
  \label{eq:bfpsi.M.far}
  \| \bfpsi^M(z)|_{\bfG^M_\far}\|_{\bfG^M_\far} \leq \frac{1}{\wt s_*} \,
\frac{C_\bbU \,\mfS_0(\alpha_\bbU)}{ \gamma \rho_0} \|\bfdelta\|_{\bfG^M}
  \quad  \text{for all } z\in \R. 
\end{equation} 
The factor in front of $\|\bfdelta\|_{\bfG^M}=\rho_0$ is small if the
``cut-off" excitation length $\ell_\mafo{excit}(s_*)=1/\wt s_*$ is small with
respect to the global scattering length $\ell_\mafo{scatt} =\rho_0/C_\bbU$. In
such a case it may be reasonable to neglect these wave vectors and solve \DHW\
on the much smaller set $\bfG^M_\Ew$ instead in all of $\bfG^M$. The error is
controlled in the following result.

\begin{theorem}[Reduction to Ewald sphere]\label{th:ReductEwald}
  Under the above assumptions consider the solutions $\bfpsi^M$ and
  $\bfpsi^M_\Ew$ of \DHW$_{\bfG^M}$ and \DHW$_{\bfG^M_\Ew}$ with initial
  condition $\bfdelta$, respectively. If $\bfG^M_\Ew$ is given by $\wt s_*$,
  then for all $z\in \R$ we have the error estimate
\begin{equation}
  \label{eq:bfpsi.M.Ew}
  \| \bfpsi^M_\Ew(z)- \bfpsi^M(z)|_{\bfG^M_\Ew}\|_{\bfG^M_\Ew} \leq
  |z|\, \frac{\pi}{\wt s_*} \frac{C_\bbU^2\;\!(\mfS_0(\alpha_\bbU){-}1) 
  \;\!\mfS_0(\alpha_\bbU)  }{\gamma^2\,\rho_0^2 } \,\|\bfdelta\|_{\bfG^M} \,. 
\end{equation}
\end{theorem}
\begin{proof} We proceed exactly as in the proof of Theorem \ref{th:ApprError}
  but the nested couple $(\bfG^M, \bfG_\gamma)$ there is replaced by the nested
  couple $(\bfG^M_\Ew , \bfG^M)$ here. The bound in Step 1 is replaced by the
  bound for $\bfpsi^M(z)|_{\bfG^M_\far}$ in \eqref{eq:bfpsi.M.far}. In Step 2
  the norm of the coupling operator can be estimated by the same constant
  $N_\mafo{cpl}$. Now 
\[
\| \bfpsi^M_\Ew(z)- \bfpsi^M(z)|_{\bfG^M_\Ew}\|_{\bfG^M_\Ew} \leq \int_0^z
N_\mafo{cpl}\| \bfpsi^M(\zeta)|_{\bfG^M_\far}\|_{\bfG^M_\far} \dd \zeta 
\]
gives the desired result. 
\end{proof} 

Estimate \eqref{eq:bfpsi.M.Ew} for $z\in [0,z_*]$ contains the main
error term 
$
\frac{z_* }{\ell_\mafo{scatt}} 
\:\frac{\ell_\mafo{excit}(\wt s_*)}{\ell_\mafo{scatt}}.
$
Because of $z_*\approx \ell_\mafo{scatt}$ it is important to have $\wt s_*$ big
enough to obtain $\ell_\mafo{excit}(\wt s_*)=1/\wt s_* \ll \ell_\mafo{scatt}$. 

However, using the fact that $|\sigma_g|\approx 2|k_0|\mathOP{dist}(g,\bbS_\Ew)$
and $|\rho_g| \approx |k_0|$ we see that the number of wave vectors in
$\bfG^M_\Ew$ is proportional to $O(\wt s_* M^{d-1})$, while the number of
wave vectors in $\bfG^M$ scales like $O(M^d)$. Thus, it is desirable to make
$\wt s_*$ even less than $1$, which means one spacing in $\DLL$ perpendicular
to $\bbS_\Ew$ (recall that
$\wt s_*$ has the physical dimension of $|k_0|$ which is an inverse
length).

\section{Special approximations}
\label{se:Approximations}

Here we discuss approximations that are commonly used in the physical
literature to interpret TEM measurements, see \cite{Jame90APTH, Degr03ICTE,
  Kirk20ACEM}.

\subsection{Free-beam approximation}
\label{su:FreeBeamApprox}

For a mathematical comparison, it is instructive to consider the trivial
approximation, where only the incoming beam is considered, i.e.\ we use
$\bfG=\{0\}$, i.e.\ the equation \DHW$_{\{0\}}$ consists
of the single ODE
\begin{equation}
  \label{eq:FreeBeam}
   \frac{\rho_0}\pi\, \dot\psi_0 =\ii (\sigma_0 {+}U_0) \psi_0, \quad \psi_0(0)=1.
\end{equation}
Using $\sigma_0=0$ we obtain the trivial solution
$\psi_0(z)= \ee^{\ii z \pi U_0/\rho_0}$ and obtain that the intensity $I_0$
remains constant: $I_0(z)=|\psi_0(z)|^2=1$. We will see that this is a reasonable
approximation for $z\in [0,z_*]$, if $z_* C_\bbU/|k_0| =z_*/\ell_\mafo{scatt}
\ll 1$, which means that 
the scattering length is small compared to the thickness $z_*$ of the
specimen.

\begin{lemma}[Free beam]\label{le:FreeBeam} 
  Choose $\gamma\in {]0,1[}$ and consider $\bfG\subset \bfG_\gamma$ with
  $0\in \bfG$. Let the solution $\bfpsi=(\psi_g)_{g\in \bfG}$ of \DHW$_\bfG$
  with initial condition $\bfpsi(0)=\bfdelta$ and let $\wh\psi_0$ be the
  solution of \eqref{eq:FreeBeam}.  Then we have the approximation errors
\begin{subequations}
  \label{eq:ErrFreeB}
\begin{align}
 \label{eq:ErrFreeB.a}
 &|\wh\psi_0(z) - \psi_0(z)| \leq \min\{ N_\mafo{cpl} |z|,\, 2\} \quad \text{with } N_\mafo{cpl} =
 \frac{\pi C_\bbU( \mfS_0(\alpha_\bbU){-}1)}{ \gamma \rho_0} , 
 \\
 \label{eq:ErrFreeB.b}
 &\| \bfpsi(z)|_{\bfG\setminus\{0\}} \|_{\bfG\setminus\{0\}} \leq \min\{ N_\mafo{cpl}
 |z|,\,1\} \|\bfdelta\|_{\bfG} \quad \text{for all } z\in \R\,.
\end{align}
\end{subequations}
\end{lemma}
\begin{proof} This result follows exactly as in Theorem \ref{th:ReductEwald},
  where we now use the a priori estimate
  $\| \bfpsi(z)|_{\bfG\setminus\{0\}} \|_{\bfG\setminus\{0\}} \leq
  \|\bfpsi(z)\|_{\bfG} = \|\bfdelta\|_{\bfG} = |\rho_0|^{1/2}$.  Then, the
  analog to \eqref{eq:bfpsi.M.Ew} gives
  $|\rho_0|^{1/2}|\wh\psi_0(z) - \psi_0(z)|\leq
  N_\mafo{cpl}|\rho_0|^{1/2}|z|$. Together with the trivial bounds
  $|\rho_0|^{1/2}|\psi_0(z)|\leq \|\bfpsi(z)\|_\bfG = |\rho_0|^{1/2}$ we arrive
  at \eqref{eq:ErrFreeB.a}.

To obtain the second equation we set
$B(z)=\bfpsi(z)|_{\bfG\setminus\{0\}} \in \mfH(\bfG {\setminus} \{0\}) $ and
apply Duhamel's principle to
$\ii R\dot B + \Sigma B- \bbU_{BB} B= \bbU_{B,\{0\}} \psi_0(z)$ and obtain
\[
  \| \bfpsi(z)|_{\bfG\setminus\{0\}} \|_{\bfG\setminus\{0\}} \leq \int_0^z
  \|\bbU_{B,\{0\}}\| \;\!\|\psi_0(\zeta)\|_{\{0\}} \dd \zeta \leq \int_0^z
  N_\mafo{cpl} |\rho_0|^{1/2} \dd \zeta = N_\mafo{cpl} |\rho_0|^{1/2}|z|.
\]  
Together with the trivial bound
$\| \bfpsi(z)|_{\bfG\setminus\{0\}} \|_{\bfG\setminus\{0\}} \leq
\|\bfpsi(z)\|_{\bfG} = |\rho_0|^{1/2}$ we find \eqref{eq:ErrFreeB.b}.
\end{proof}

Thus, this trivial result provides a rigorous quantitative estimate for the
obvious fact that the incoming beam stays undisturbed only if the thickness $z_*$
of the specimen is significantly shorter than the scattering length
$|k_0|/C_\bbU$, i.e.\ $N_\mafo{cpl}z_*\ll 1$.

\subsection{Approximation via the lowest-order Laue zone}
\label{su:LOLZ}

The lowest-order Laue zone (LOLZ) is defined if the wave vectors in the tangent
plane $\bfT_{k_0}:=\bigset{\eta \in \R^d}{ \eta\cdot k_0=0}$ to the Ewald sphere
$\bbS_\Ew$ at $g=0$ form a lattice of dimension $d-1$. Denoting by $\kappa_*$
the minimal distance between different points in $\DLL$ we define
\[
\bfG_\mafo{LOLZ} := \bigset{g \in \DLL\cap \bfT_{k_0} }{ \mathOP{dist}( g, \bbS_\Ew)
  \leq \kappa_*/2}, 
\] 
see Figure \ref{fig:Ewald} for an illustration.  Because the Ewald sphere can
be approximated by the parabola $g_z= - |g_x|^2/(2|k_0|)$, the set
$\bfG_\mafo{LOLZ}$ is contained in a circle of radius
$M:=m_*|k_0|^{1/2}$ inside $\bfT$, where $m_*=\kappa_*^{1/2}$. 
(To include higher-order Laue zones up to order $n$ one chooses
$m_*=\big((2n{+}1)\kappa_*\big)^{1/2}$.)

This observation allows us to assess the approximation error for the solution
$\bfpsi^\mafo{LOLZ} $ that is obtained by solving the DHW equations on
$\mfH(\bfG_\mafo{LOLZ})$. For this, we first use Theorem \ref{th:ApprError} to
reduce to the set $\bfG^M$ with $M=m_*|k_0|^{1/2}$, and second we reduce to the
$ \bfG_\mafo{LOLZ} = \bfG^M_\Ew$ using Theorem \ref{th:ReductEwald} with a
suitable $\wt s_* \sim \kappa_*$. In the following result we give up the exact
formulas for the constants in the error estimate. In particular, we will drop
the dependence on $\alpha_\bbU$, which we consider to be fixed. However, we
keep the dependence on $|k_0|$ and $C_\bbU$ to the influence of the energy
and the scattering.  To achieve formulas with correct physical dimensions we
sometimes use the length scale $\alpha_*$, which could be chosen as the lattice
constant of $\Lambda$, as $1/\kappa_*$, or simply $\alpha_\bbU$.  We will use
generic, dimensionless constants $N$ and $N_j$ that may change from line to
line and will depend on $\alpha_*$ and $\alpha_\bbU$, but do not depend on
$|k_0|$ and $C_\bbU$.

\begin{theorem}[LOLZ approximation] \label{th:psiLOLZ}
  Consider the solution $\bfpsi^\gamma$ of \DHW$_{\bfG_\gamma}$ with
  $\gamma=1/2$ and the solution $\bfpsi^\mafo{LOLZ}$ of
  \DHW$_{\bfG_\mafo{LOLZ}}$ for the initial condition $\bfdelta$. Given a
  constant $N_0>0$ there exists constants $k_*$ and $N_1$ such that the
  following holds:
\begin{subequations}
  \label{eq:Appr.LOLZ}
\begin{align}
  \label{eq:z*.cond}
&\text{If }|k_0|\geq k_* \text{ and } z_* \leq  N_0 |\alpha_* k_0|^{1/3}  
 \frac{|k_0|}{C_\bbU}, \text{ then for all } z\in [0,z_*] \text{ we have}
\\
  \label{eq:Appr.LOLZ.b}
&\| \bfpsi^\mafo{LOLZ}(z)- \bfpsi^\gamma(z)|_{\bfG_\mafo{LOLZ}} 
 \|_{\bfG_\mafo{LOLZ}} \leq N_1 \Big( \frac1{|\alpha_* k_0|^2} +
 \frac{\alpha_* C_\bbU^2}{|k_0|^2}  \,z_*\Big)\|\bfdelta\|_\bfG\,.
\end{align}
\end{subequations}
\end{theorem}
\begin{proof} \STEP{Step 1. Reduction to $\bfG^M$.} Using  $M=m_*|k_0|^{1/2}$ and
  $|k_0|\geq k_*$ we have $\bfG^M\subset \bfG_\gamma$, and 
  Theorem  \ref{th:ApprError} with $\alpha= \alpha_\bbU/2$ provides the error estimate 
\[ 
\| \bfpsi^M(z) - \bfpsi^\gamma(z)|_{\bfG^M}\|_{\bfG^M} \leq N_2\,\ee^{N_3C_\bbU
  z/|k_0| - N_4|\alpha_* k_0|^{1/2}} \|\bfdelta\|_\bfG\, . 
\]

\STEP{Step 2. Reduction to $\bfG_\mafo{LOLZ}$.} The theory in Section
\ref{su:Averaging} reduces to the Ewald sphere. In particular, because of our given
radius $M=m_*|k_0|^{1/2}$ the set $\bfG^M_\Ew$ exactly equals
$\bfG_\mafo{LOLZ}$ if we choose the cut-off value $\wt s_*$ suitably. 

For this we have to identify the smallest value of $|\sigma_g/\rho_g|$ in
$\bfG^M{\setminus} \bfG_\mafo{LOLZ}$. Because $\rho_g=k_0\cdot \bfnu
+ O(|k_0|^{1/2}) $ in $\bfG^M$, it suffices to minimize $|\sigma_g|$ in
$\bfG^M_\far=\bfG^M{\setminus} \bfG_\mafo{LOLZ}$, or simply minimize
the distance to $\bbS_\Ew$. Hence, the points in the interior of the 
Ewald balls in the lattice layer right below $\bfG_\mafo{LOLZ} \subset \bfT$ are most
critical. All of them have distance $\kappa_*$ to $\bfT$ and thus their
distance to the Ewald sphere is bigger or equal $\kappa_* - M^2/(2|k_0|) =
\kappa_*/2>0$. 

From this, for $g \in \bfG^M_\far$ one has
$|s_g|\geq \frac{\kappa_*}2 \;\!|k_0|$, and with $\rho_0\approx |k_0|$ we are
able to apply Theorem \ref{th:ReductEwald} with $\wt\sigma_*= \kappa_*/3$,
which is independent of $|k_0|$ and $\C_\bbU$. With this we conclude
$\| \bfpsi^\mafo{LOLZ} (z) -\bfpsi^M(z) |_{\bfG_\mafo{LOLZ}}
\|_{\bfG_\mafo{LOLZ}} \leq |z| N_4 C_\bbU^2 \|\bfdelta\|_\bfG /|k_0|^2$ for all
$z\in \R$.

\STEP{Step 3. Combined estimate.} We observe that the second relation in
\eqref{eq:z*.cond} allows us to simplify the estimate in Step 1. For $z\in
[0,z_*]$ the exponent can be estimated via 
\[
N_3C_\bbU z/|k_0| - N_4|\alpha_* k_0|^{1/2} \leq N_3 N_0|\alpha_*k_0|^{1/3} -
N_4|\alpha_* k_0|^{1/2} \leq N_5 - \tfrac{N_4}2|\alpha_* k_0|^{1/2}.
\]
Now, the final result follows $\ee^{-N_6|k_0|^{1/2}}\leq N_7/|k_0|^2$ and the
previous two steps.
\end{proof}

\subsection{Two-beam approximation and beating}
\label{su:TwoBeam}

The most simple nontrivial approximation is obtained by assuming that the
incoming beam at $g=0$ interacts mainly with one other wave vector $\whg$.
The energy exchange between $\psi_0$ and $\psi_\whg$ is called
beating and occurs on a well controllable length scale. Thus, it can
be used effectively for generating contrast in microscopy, see
\cite{Darw14TXRR12} or 
\cite[Sec.\,4]{MNSTK20NSTI}.

The theory is often explained by the following two-equation approximation of
DHW with $\bfG=\{0,\whg\}$, but even though it turns out that this model
predicts nicely certain qualitative features it is not accurate enough for
quantitative predictions. For a typical microscopical experiment, one chooses
$k_0$ such that $g=0$ and $g=\whg$ are the only two wave vectors on the Ewald
sphere:
\begin{equation}
  \label{eq:TwoBeamAss}
  \sigma_0 = \sigma_\whg=0 \quad \text{and} \quad \rho_0 = \bfnu\cdot k_0=\rho_\whg.
\end{equation}
Assuming $\whg=(0,n,0) \in \bfG$ with a small integer $n$ this can be achieved
by setting $k_0=(\theta,{-}n/2,k)$ with $k\approx |k_0|\gg1$ and $|\theta|<1$, see
Figure \ref{fig:TwoBeam}.
\begin{figure}
\begin{minipage}{0.6\textwidth}
\centering
  \begin{tikzpicture}[scale=0.8]
    \begin{scope}
      \clip (-2.04,-3.35) rectangle (6.04,1.0);
      \fill [color=gray!10]  (-5.5,-3.5) rectangle (6.5,2.5);
       \foreach \i in {-18,...,20}
         \foreach \j in {-9,...,6}
          \fill (0.4*\i+0.1*\j , 0.4*\j) circle(0.06);
       \draw[ultra thick, color=blue] (2,-6) circle (6.325);

  \fill (0,0) circle (0.14);
  \fill (4,0) circle (0.14) node[above=0.13, fill =gray!10]{$\whg$};

  \node[color=blue] at (6.5,-1.3){$\bbS_\Ew$};
  \draw[color=red, ultra thick,-{Latex[length=5mm]}] (2,-6) -- (0,0)
                               node [pos=0.75, right=0.2, fill=gray!10]{$k_0$}  ;
  \draw[color=red, ultra thick,-{Latex[length=5mm]}] (2,-6) -- (4,0)
                               node [pos=0.75, right=0.2, fill=gray!10]{$k_0{+}\whg$}  ;

    \end{scope}
  \draw[thick,->] (-2.2,0) -- (6.5,0) node[right]{$g_y$} ;
  \draw[thick,->] (0,-3.5) -- (0,1.3) node[right]{$g_z$}; 
  \end{tikzpicture}
\end{minipage}
\begin{minipage}[c]{0.33\textwidth}
  \caption{A typical setup for the two-beam conditions: $g=0$ and $g=\whg$ are
    the only two points in $\bbS_\Ew\cap \DLL$. }
\label{fig:TwoBeam}
\end{minipage}
\end{figure} 
Then, the two-equation model  for $g=0$ and $g=\whg$ reads: 
\begin{equation}
  \label{eq:TwoBeamMod}
  \frac{\rho_0}{\pi}\, \dot \psi_0 = \ii\big(\sigma_0 \psi_0 + U_0 \psi_0 + 
  \ol U_{\whg} \psi_\whg\big) , \quad 
\frac{\rho_\whg}{\pi} \dot \psi_{\whg}= \ii \big( \sigma_\whg \psi_\whg + U_\whg \psi_0
+ U_0 \psi_\whg\big)\,.
\end{equation}
This complex two-dimensional and real four-dimensional system can be
solved explicitly leading to quasi-periodic motions with the frequencies 
$\omega_{1,2} = \pi(U_0\pm |U_\whg|)/\rho_0$, where we used
\eqref{eq:TwoBeamAss} to simplify the general expression. 

Recalling the wave-flux conservation from Proposition
\ref{pr:ExUniFlux} we obtain the relation 
$\rho_0 |\psi_0(t)|^2 +\rho_\whg |\psi_\whg(t)|^2 = \rho_0 $ by using
the initial condition $\bfpsi(0)=\bfdelta$. A direct, but lengthy
calculation gives 
\begin{equation}
  \label{eq:ApproxBeat}
  I_0(z):= |\psi_0(z)|^2 = \big( \cos( \frac{\pi|U_\whg|}{\rho_0}\!\; z)\big)^2 \ \text{
    and } \ I_\whg(z):=|\psi_\whg(t)|^2 = \big( \sin( \frac{\pi|U_\whg|} 
   {\rho_0}\!\; z)\big)^2,
\end{equation}
which clearly displays the energy exchange, also called beating.

We do not give a proof for the validity of the two-beam approximation, but
rather address its limitations. However, we refer to the systematic-row
approximation in the next section, which includes the two-beam approximation
as a special case. To see the limitation we simply argue that having the beams
in $g=0$ and $g=\whg=(0,n,0)$, we also have scattering from $g=0$ to the neighbors
$(0,j,0)$. This scattering must be small if the two-beam approximation should
be good. The smallness can happen if one of the following reasons occurs: (i)
the scattering coefficient $U_{(0,j,0)}$ is $0$ or very small or (ii) the
excitation error $s_{(0,j,0)}$ is already big. The first case may indeed occur,
e.g.\ for symmetry reasons, however, because beating needs a reasonably large 
$U_{\whg}=U_{(0,n,0)}$ we also have that $U_{(0,-n,0)}= \ol U_\whg$  is
reasonably large. Hence, in this case only the reason (ii) can be valid, i.e.\
we need $|s_{(0,-n,0)}| \gg \pi|U_{(0,-n,0)}|/\rho_{(0,-n,0)} \approx
3|U_\whg|/|k_0| $. Using $\sigma_0=\sigma_\whg=0$, the excitation error has the
expansion $s_{(0,j,0)} \approx (nj{-}j^2)/(n|k_0|)$, which leads with $j=-n$ to the
condition $3|U_\whg| \ll |n|$, which is not easily satisfied.    
 
Indeed, in \cite{MNSTK20NSTI} TEM imaging is done under two-beam conditions for
$\whg =(0,4,0)$, where $U_{(0,j,0)}=0$ for odd $j$ and
$U_{(0,2,0)} \approx - 0.05\;\!U_{(0,4,0)}$.  In particular, $j=4$ was chosen,
because it gives the biggest value for $|U_{(0,j,0)}|$ for $j\neq 0$.
Nevertheless, it was necessary to base the analysis of the TEM images in the
solution $\bfpsi$ of \DHW$_\bfG$ for $\bfG= \bfG^M_\Ew$ obtained via the
software package pyTEM. The simple usage of the approximations in
\eqref{eq:ApproxBeat} would not be sufficient.

We will see in Section \ref{se:Applications} that even in simple examples
the two-beam approximation is only a very rough approximation, see e.g.\ Figure
\ref{fig:solutions}.

\subsection{Systematic-row approximation}
\label{su:SystematicRow}
 
We choose 
\[
\bfG= \bigset{ n \,g_* }{n_{\min} \le n \le n_{\max} },
\] 
where $g_*$ is small and almost perpendicular to $k_0$, such that the convex
hull of the set $\bfG$ is roughly tangent to the Ewald sphere $\bbS_\Ew$.  Of
course, this set should coincide with $\bfG^M_\Ew$ of Section
\ref{su:Averaging}, which can be achieved by choosing an appropriate $k_0$.  In
particular, the case of two-beam conditions of Section \ref{su:TwoBeam} can always be
seen as embedded into a systematic-row case.

Indeed, consider the simple dual lattice $\DLL=\Z^3$ and choose
$k_0=(k_x,0,k_z)$ where now $1 \ll k_x \ll k_z\approx |k_0|$, i.e.\
the incoming wave has a small, but nontrivial angle to the normal $\bfnu$ of the
specimen. Assuming $k_x = c_*|k_0|^{2/3}$ and considering only $g\in
\bfG^M=B^M(0)\cap \Z^3$ with $M=|k_0|^{1/4}$ we see that 
\[
\sigma_g =|k_0|^2 -|k_0{+}g|^2 \approx - g_x^2 -g_y^2-g_z^2
-2c_*|k_0|^{2/3}g_x-2|k_0|g_z   
\]
can only take values smaller than $O(|k_0|^{1/2})$ if the wave vectors satisfy
$g_x=g_z=0$, i.e.\ $g= n(0,1,0)$ with $|n| \leq |k_0|^{1/4}$, which is a finite
row of wave vectors.  

Moreover, in $\bfG^M$ we have $\rho_g = (k_0{+}g)\cdot \bfnu= |k_0|+
O(|k_0|^{1/2})$ and conclude   
\begin{align*}
&\bfG_\mafo{syrow}:=\bigset{ (0,n,0) }{ |n|\leq |k_0|^{1/4}} 
 = \bfG^M_\Ew:=\bigset{g\in \bfG^M}{ |s_g|/(2\rho_g)< \wh\sigma_*}\\
&\qquad\qquad \text{with } M=\kappa_*^{3/4}|k_0|^{1/4} \text{ and }
\wh\sigma_*=\kappa_*^{3/2}|k_0|^{-1/2}. 
\end{align*}
Thus, as for the case of the lowest-order Laue zone we obtain an error estimate. 

\begin{theorem}[Systematic-row approximation]\label{th;:SystRow}
Under the above assumptions consider the solutions
$\bfpsi^\gamma$ and $\bfpsi_\mafo{syrow}$ of \DHW$_{\bfG_\gamma}$ with
$\gamma=1/2$ and \DHW$_{\bfG_\mafo{syrow}}$, respectively. Then, for all
$N_0$ there exists  $k_*$ and 
$N_1$ such that  the following holds: 
\begin{subequations}
  \label{eq:SystRow}
\begin{align}
  \label{eq:z*.SystRow}
&\text{If } |k_0|\geq k_* \text{ and } 
 z_* \leq  N_0 |\alpha_* k_0|^{1/5} \frac{|k_0|}{C_\bbU}, 
\text{ then for all } z\in [0,z_*] \text{ we have}
\\
  \label{eq:SystRow.b}
&\| \bfpsi_\mafo{syrow}(z)- \bfpsi^\gamma(z)|_{\bfG_\mafo{syrow}}
 \|_{\bfG_\mafo{syrow}} \leq N_1 \Big( \frac1{|\alpha_* k_0|^2} +
 \frac{\alpha_*^{3/2} C_\bbU^2}{|k_0|^{3/2} } \,z_*\Big)\|\bfdelta\|_\bfG\,.
\end{align}
\end{subequations}
\end{theorem}
\begin{proof} \STEP{Step 1. Reduction to $\bfG^M$.}  
 Using  $M=\kappa^{3/4}|k_0|^{1/4}$ and
  $|k_0|\geq k_*$ we have $\bfG^M\subset \bfG_\gamma=\bfG_{1/2}$, and 
  Theorem  \ref{th:ApprError} with $\alpha= \alpha_\bbU/2$ provides the error estimate 
\[ 
\| \bfpsi^M(z) - \bfpsi^\gamma(z)|_{\bfG^M}\|_{\bfG^M} \leq N_2\,\ee^{N_3C_\bbU
  z/|k_0| - N_4|\alpha_* k_0|^{1/4}} \|\bfdelta\|_\bfG\, . 
\]

\STEP{Step 2. Reduction to $\bfG_\mafo{syrow}$.} Applying Theorem \ref{th:ReductEwald}
with $\wt s_* = \kappa_*^{3/2}|k_0|^{-1/2}$ we obtain the error bound 
$
  \| \bfpsi_\mafo{syrow}(z)-
  \bfpsi^M(z)|_{\bfG_\mafo{syrow}}\|_{\bfG_\mafo{syrow}}  \leq
  N_5 \frac{z_*}{\kappa_*^{3/2}} \frac{C_\bbU^2 }{|k_0|^{3/2}}
  \,\|\bfdelta\|_{\bfG^M}.
$

\STEP{Step 3. Combined estimate.} We conclude as in Step 3 of the proof of
Theorem \ref{th:psiLOLZ}.
\end{proof} 

In contrast to the cut-off choice $\wt\sigma_*\sim 1$ for the Laue-zone approximation
we have now chosen $\wt s_* \sim |k_0|^{-1/2}$. This reduces the number of
points in the systematic-row 
approximation, i.e.\ the number of coupled ODEs to be solved is proportionally
$|k_0|^{1/4}$, whereas for the Laue-zone approximation, the number of ODEs is
proportional to $|k_0|$.  However, the gain in computation power is accompanied
by a loss of accuracy and a smaller domain of applicability, see
Figure~\ref{fig:CompareTable}.

\begin{figure}
\centerline{\small
\begin{tabular}{|c||c|c|} \hline 
approximation  & Laue zone& systematic row\\ \hline
$(M,\wt\sigma)$\rule[-0.7em]{0em}{1.9em}& $(|k_0|^{1/2}, \,1\,)$& $(|k_0|^{1/4}, |k_0|^{-1/2})$  \\ \hline
number of points\rule[-0.7em]{0em}{1.9em}&  $|k_0|$  & $|k_0|^{1/4}$ \\ \hline
\!\!thickness restriction\rule[-0.7em]{0em}{1.9em}\!\! 
  &\!\! $z_*\leq N_0 |k_0|^{1/3}\ell_\mafo{scatt}$ \!\!
    &\!\! $z_*\leq N_0 |k_0|^{1/5}\ell_\mafo{scatt}$\!\!   \\ \hline
first error term\rule[-0.7em]{0em}{1.9em}&  $|\alpha_* k_0|^{-2}$ &  $|\alpha_* k_0|^{-2}$  \\ \hline 
second error term\rule[-1.3em]{0em}{3.1em}& $\dfrac{\alpha_* z_*}{\ell_\mafo{scatt}^2}$ & 
          $\dfrac{\alpha^{3/2}_*|k_0|^{1/2} z_*}{\ell_\mafo{scatt}^2}$
         \\ \hline
\end{tabular}
\ 
\begin{minipage}{0.26\textwidth}
\caption{Comparison of the Laue approximation in Section \ref{su:LOLZ}) and
  the systematic-row approximation.}
  \label{fig:CompareTable}
\end{minipage}
}
\end{figure}

\section{Simulations for TEM experiments}
\label{se:Applications}

Here we provide a numerical example of the DHW equations, compare the
solutions for different choices of the wave-vector set $\bfG$, and relate the
observed errors with the mathematical bounds established above.

To make our simulations close to values in real TEM, we choose
the lattice constant $a_0=0.56503\,\mafo{nm}$ of a GaAs crystal and the
specimen thickness $z_*= 200a_0 = 113.006 \,\mafo{nm} $.  At the TEM typical
acceleration voltage 400\,kV, the wave length is
$\lambda=1/|k_0|=1.644\,\mafo{pm}$, which in normalized dimensions is
$\lambda = 0.00294 a_0$.  This gives us a wave vector of
$|k_0|= 608.293 \mafo{nm}^{-1}$.  The full system to consist of 30
beams:
\[
\bfG= \bigset{ g = \frac{4}{a_0} (\tilde{g}_y,\tilde{g}_z) }{\tilde{g}_y\in
  \{-2,\dots,3\} \text{ and } \tilde{g}_z\in \{-2,\dots,2\}}. 
\] 
One would expect to use a beam list of the form
$\bigset{ g = \frac{1}{a_0} (\tilde{g}_y,\tilde{g}_z) }{\tilde{g}_y,\tilde{g}_z
  \in \mathbb{Z}}$.  But for GaAs the scattering potential has significant
contributions $U_g$ only for beams of the form in $\bfG$, while the other $U_g$
are small or even 0, see Figure \ref{th:WeightedEstim}. This is due to the
face-centered cubic lattice of the crystal and the properties of the Ga and As
atomic form factors.  Therefore, we restrict our beam list to that case in our
example.

For the potential we use $U_{(0,0)}=10 \,\mafo{nm^{-2}}$,
$U_{(\pm1,0)}=U_{(0,\pm1)}=3 \,\mafo {nm^{-2}}$, and
$U_{(\pm1,\pm1)} =U_{(\pm1,\mp1)} =2 \,\mafo {nm^{-2}}$ and $U_{\tilde{g}} = 0$
for the rest.  We consider strong beam excitation
$(\tilde{g}_y,\tilde{g}_z)=(1,0)$ corresponding to $g=\frac{1}{a_0} (4,0)$ and
$k_0=(-2/a_0, 608.293)$.


We first solve the DHW equations for $\bfG$ with 30 beams as a reference
solution. Note that in 2D there
is no distinction between Laue zone and systematic-row approximation. 
Figure~\ref{fig:exc_error} displays the excitation errors $s_g$:
In the middle row, which corresponds to the points close to the Ewald
sphere, the entries have a modulus that is more than a factor of 10 smaller than in
the rows above and below. We have a zero excitation error at $(0,0)$ and
$(1,0)$, due to our strong beam excitation condition. 
\begin{figure}
\centerline{\begin{tabular}{|c||c|c|c|c|c|c|} \hline 
$\tilde{g}_z \,\backslash \, \tilde{g}_y$ 
   & -2      & -1      &    0 &  1 & 2 & 3 \\ \hline\hline
-2 &  -14.07 &  -14.24 & -14.33 & -14.33  & -14.24 & -14.07  \\ \hline
-1 & -6.87 & -7.04  & -7.12 & -7.12  &  -7.04 &  -6.87 \\ \hline
0 &  0.25 &  0.08 &  0 &  0 &  0.08 & 0.25 \\ \hline
1 & 7.28 & 7.12 & 7.04 & 7.04 & 7.12 & 7.28 \\ \hline
2 & 14.24 & 14.09 & 14.00 & 14.00 &  14.08 & 14.24 \\ \hline
\end{tabular}}
\caption{Excitation errors $s_g=\sigma_g/(2\rho_g)$ for every point $g=
  \frac4{a_0} (\tilde g_y,\tilde g_z) \in \bfG$. The
  middle row corresponds to beams of the systematic-row approximation. 
  \label{fig:exc_error}}
\end{figure} 

\begin{figure}
\begin{minipage}{0.6\textwidth}
\begin{tikzpicture}
\node at (0,0){
  \includegraphics[width=0.8\textwidth, trim= 30 44 150 90, clip=true]
    {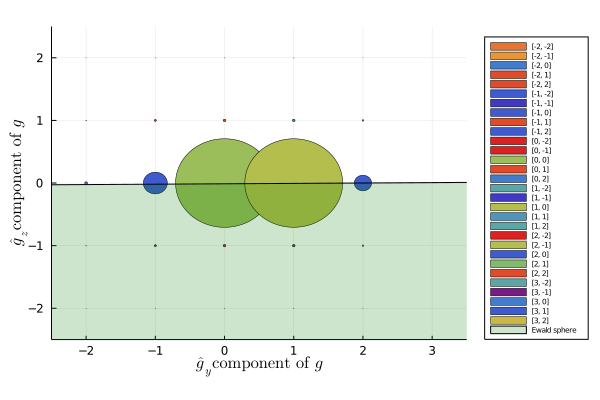} }; 
 \node  at (1,-2.4) {component $\tilde g_y$ of $g\in \bfG$};
 \node[rotate=90]  at (-3.5,0.02) {component $\tilde g_x$ of $g\in \bfG$};
\end{tikzpicture}
\end{minipage}
\hfill
\begin{minipage}{0.39\textwidth}
\caption{The radius of the circles correspond to the amplitudes $|\psi_g(z_*)|$.  
  Close to the Ewald sphere (boundary of light green area) the excitation
  errors are significantly smaller and the 
  amplitudes are much larger. All simulations are done in
  Julia.}
\label{fig:ewald.ampl}
\end{minipage}
\end{figure}

Figure \ref{fig:ewald.ampl} shows that the amplitudes of the numerical solutions are
related to the excitation errors. For each beam $g$, we plot a circle with
center $(\tilde{g}_y,\tilde{g}_z)$ and radius proportional to $|\psi_g(z_*)|$.  
We see that near the Ewald sphere, where the
excitation error is small, the amplitude is significantly higher.  It becomes 
obvious that there are four main modes, corresponding to the beams $(-1,0)$,
$(0,0)$, $(1,0)$, and $(2,0)$.

\begin{figure}
\raisebox{0.13\textwidth}{$|\psi_g|$}%
\includegraphics[height=0.23\textwidth, trim=135 80 120 70, clip=true]{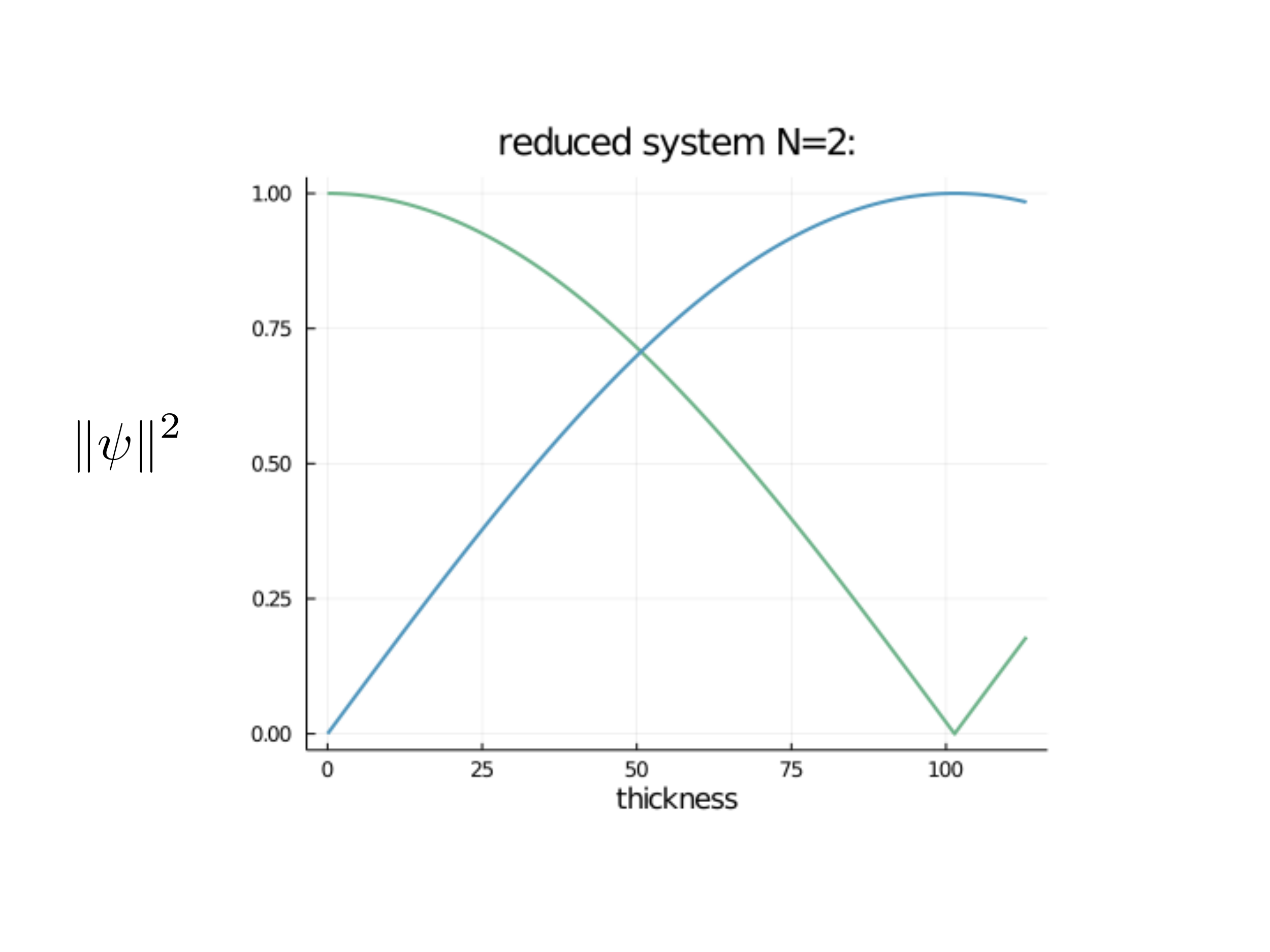}
\hfill
\includegraphics[height=0.23\textwidth, trim=130 80 120 70, clip=true]{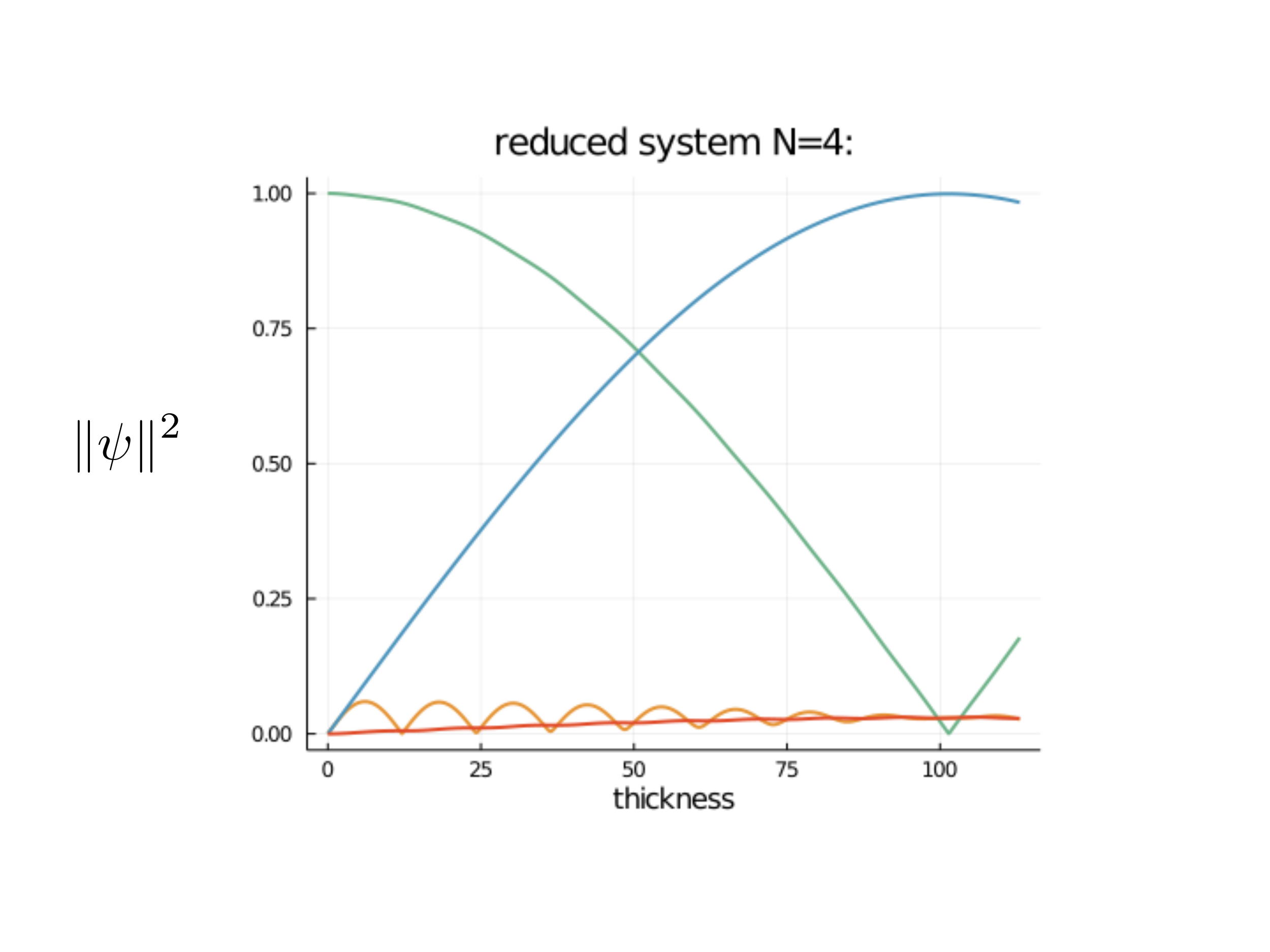}
\hfill
\includegraphics[height=0.23\textwidth, trim=130 80 10 70,
clip=true]{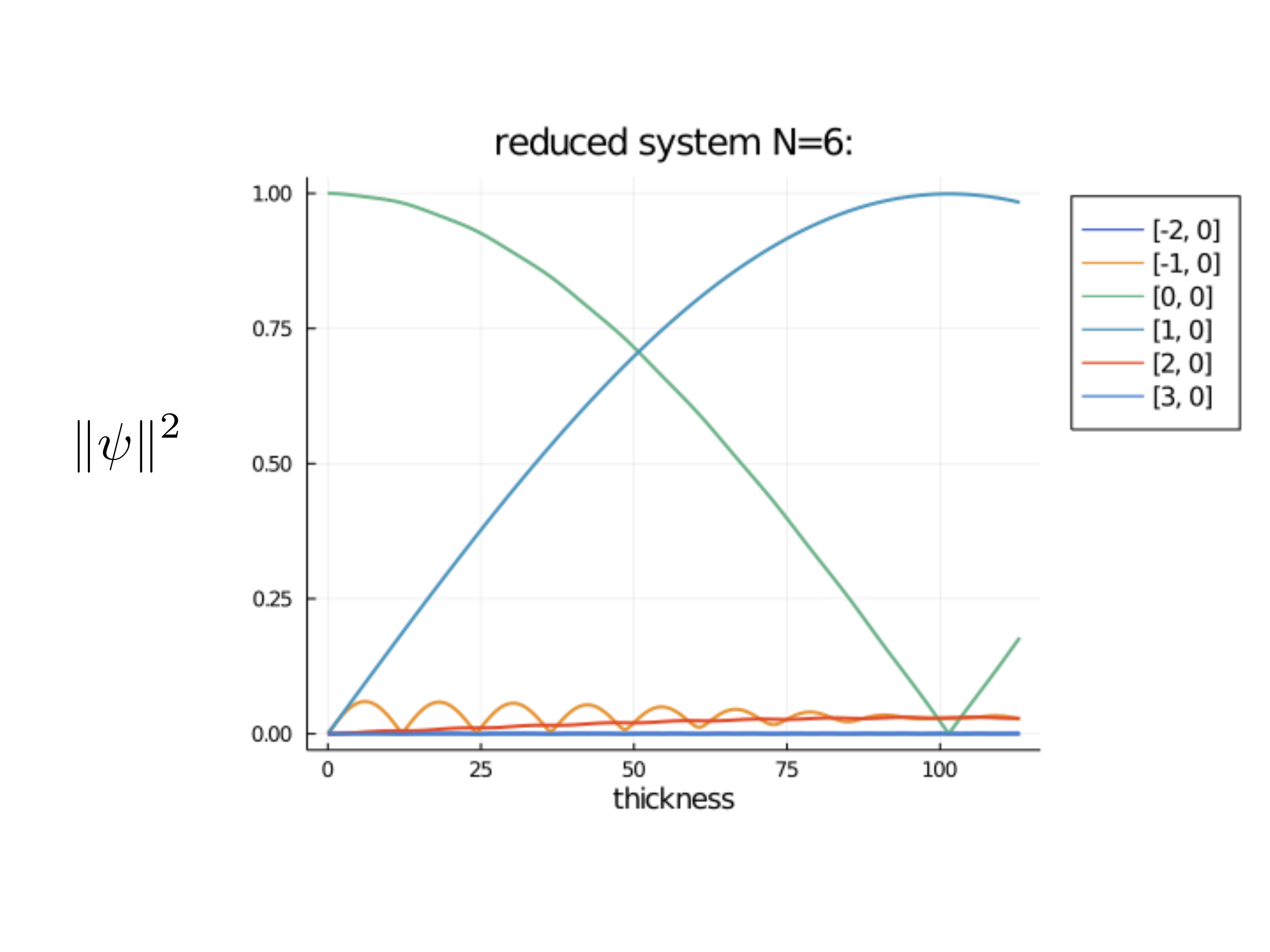}

\caption{Beam amplitudes $|\psi_g(z)|$ of solutions for the three choices
  $\bfG_1$, $\bfG_2$, and $\bfG_3$. The same beating of the two main modes
  is observed in all the cases.} 
\label{sys_row}

\end{figure}

Next, we reduce the beam list $\bfG$ to observe how the errors of the
solutions change.  We create three sets corresponding to the systematic-row
approximation:
\begin{align*}
  & \bfG_1= \bigset{ g = \frac{4}{a_0}(\tilde{g}_y,\tilde{g}_z)
  }{\tilde{g}_y\in \{0,1\} \text{ and } \tilde{g}_z\in \{0\}}, \\[-0.1em] 
  & \bfG_2= \bigset{ g = \frac{4}{a_0}(\tilde{g}_y,\tilde{g}_z)
  }{\tilde{g}_y\in \{-1,\cdots,2\} \text{ and } \tilde{g}_z\in \{0\}}, \\[-0.1em]  
  & \bfG_3= \bigset{ g = \frac{4}{a_0}(\tilde{g}_y,\tilde{g}_z)
  }{\tilde{g}_y\in \{-2,\cdots,3\} \text{ and } \tilde{g}_z\in \{0\}}, 
\end{align*}
where the set $\bfG_1$ corresponds to the two-beam case, shown in
Figure~\ref{sys_row}. For comparison, we also create a set including beams
above and below the Ewald sphere
\[
  \bfG_4= \bigset{ g = \frac{4}{a_0}(\tilde{g}_y,\tilde{g}_z) }{\tilde{g}_y\in
    \{-2,\cdots,3\} \text{ and } \tilde{g}_z\in \{-1,\cdots,1\}}.
\]

From Figure \ref{sys_row} we have a first qualitative comparison for the
systematic-row cases.  We see that the qualitative features, meaning the
beating and the two main modes, namely $(0,0)$ and $(1,0)$, are captured in
every case. The two beam case however fails to capture the other two main
modes, for $(-1,0)$ and $(2,0)$.

To obtain a quantitative comparison of the different models we show the
numerical values of $\psi_{(1,0)}(z_*)$ and $\psi_{(2,0)}(z_*)$ in Figure
\ref{fig:solutions}. As a first observation we see that the two-beam case has
only one significant digit correct, making it a very rough approximation.
Similar limitations of the two-beam approximation were observed in
\cite{SchSta93,WuSch19TEMD} when doing simulations with \texttt{CUFOUR}.

Moving to $\bfG_2$ with four beams gives an accuracy of 4 significant digits,
while increasing the size of the systematic-row approximation further does not
bring higher accuracy as $\bfG_3$ with six beams still has only four
significant digits.  The accuracy of the solutions can only be improved by
going beyond the systematic-row approximation (see Section \ref{su:SystematicRow}).
Indeed, we obtain 7 significant digits by adding the layer above and below
the Ewald sphere in the $\bfG_4$ system, which has 18 beams.

\begin{figure}[h]
\newcommand{\CO}[1]{\underline{#1}} 
\newcommand{\IG}[1]{}    
\centerline{\small
\begin{tabular}{ |c|c|c| c| }\hline
\!System\! & $(0,0)$ mode & $(1,0)$ mode &\!digits\!\\ \hline  \hline
$\bfG_1$&$-\CO{0.16}153606468\IG{515436} - \CO{0.07}740830300\IG{121596}\,\ii$ 
                   &$\CO{0.4}2515771142\IG{594594} -\CO{0.8}8721717658\IG{78514}\,\ii$
&1 \\ \hline
$\bfG_2$&$-\CO{0.1644}6909478\IG{421892} - \CO{0.0676}6454587\IG{452628}\,\ii$
                   &$\CO{0.3779}0257701\IG{980984} - \CO{0.9077}5029000\IG{83523}\,\ii$
&4\\ \hline
$\bfG_3$&$-\CO{0.1644}5260546\IG{214072} - \CO{0.067648}75833\IG{09119}\,\ii$
                   &  $\CO{0.377}89496977\IG{714393} - \CO{0.9077}5362575\IG{80648}\,\ii$
&4 \\ \hline
$\bfG_4$&$-\CO{0.1644425}2690\IG{38549} - \CO{0.0676480}8597\IG{630736}\,\ii$
                   & $\CO{0.3779141}0830\IG{623947} - \CO{0.9077468}3865\IG{88975}\,\ii$
&7  \\ \hline
$\bfG$ &$-0.16444251537\IG{363222} - 0.06764807576\IG{043655}\,\ii$
                   & $0.37791412093\IG{80359} - 0.90774682391\IG{440392}\,\ii$
&--- \\ \hline
\end{tabular} }

\caption{Comparison of solutions for $\tilde{g}=(0,0)$ and $\tilde{g}=(1,0)$. The
  underlined decimals indicate which numbers are already correct (up to
  rounding) with respect to the last line, i.e. we take the  $\bfG$
  system as reference.\label{fig:solutions}\vspace*{-0.5em}}
\end{figure} 

For comparing numerical errors with the mathematical error bounds in Figure
\ref{fig:CompareTable}, we observe that the scattering length is
$ \ell_\mafo{scatt} = \frac{|k_0|}{|C_{U}|} = 60.83\, \mafo{nm} \approx
z_*/2$. Choosing $\alpha_*=a_0=0.565\,\mafo{nm}$ and using
$|k_0|=0.608 \cdot 10^{12}\,\mafo{m}^{-1}$, we find the error terms\vspace{-0.2em}
\[
|\alpha_{*} k_0|^{-2}= 0.0000084647989 \lll 1 \quad \text{and} \quad \frac{\alpha_{*}
  z_{*}}{ \ell_{\mafo{scatt}}^2}=0.01735570 \ll 1,\vspace*{-0.2em}
\] 
which are indeed small for the chosen setup.\smallskip

For many practical purposes, like the simulation of TEM images with pyTEM as
in \cite{Nier19pyTEM} (see Figure~\ref{fig:tem.image}) an accuracy of 4
significant digits is certainly good enough. However, for other applications
higher accuracy may be needed, e.g.\ for detecting phase differences for beams of
low amplitudes like in electron holography, see e.g.~\cite{Lich13EHPM}. 

In fact, the software pyTEM creates a beam list $\bfG$ in the following way. 
It first restricts to the LOLZ or systematic-row approximation by setting
$g_z=0$. Next, a minimum for $|U_g|$ is chosen to restrict to the sublattice
generated by those $g$ with $|U_g|\geq u_\text{min}$. For instance, the
coefficients displayed in Figure \ref{fig:uscatt}(b) lead to the sublattice
$\bigset{g=\frac1{a_0}(0,2m,0)}{m\in \Z}$. Finally, a maximum value
$\wt s_*$ is chosen for the excitation error $s_g$, which leads to a final
systematic row approximation with 12 beams with $m\in \{-5,\ldots,6 \}$. 
Thus, it covers the same range as our set $\bfG_3$. \smallskip

\paragraph*{Acknowledgments.} This research has been partially funded by
Deutsche For\-schungs\-ge\-mein\-schaft (DFG) through the Berlin Mathematics
Research Center MATH+ (EXC-2046/1, project ID: 390685689) via the Berlin
Mathematical School, the project \hbox{EF3-1:} \emph{Model-based geometry
  reconstruction from TEM images}, and the project AA2-5 \emph{Data-driven
  electronic-structure calculations for nanostructures}. The authors are
grateful to Tore Niermann, TU Berlin, for helpful and stimulating
discussions. 

\footnotesize


%

\end{document}